\documentclass[sigconf, nonacm, dvipsnames]{acmart}

\newcommand\vldbavailabilityurl{https://github.com/vincenzo-gulisano/aggspes}

\renewcommand\footnotetextcopyrightpermission[1]{} 
\usepackage{tikz}
\usetikzlibrary{decorations.pathreplacing}
\usetikzlibrary{math} 
\usetikzlibrary{tikzmark}
\usetikzlibrary{calc}
\usepackage{nicefrac}

\usepackage{framed,enumitem}
\usepackage{booktabs} 
\usepackage[utf8]{inputenc}
\usepackage{tcolorbox}

\usepackage{listings}
\usepackage{longtable}
\usepackage{multirow}
\usepackage{graphicx}
\usepackage{subcaption}
\usepackage[noend]{algpseudocode}
\usepackage{soul}
\usepackage{comment}

\usepackage{amsthm}
\theoremstyle{definition}
\newtheorem{definition}{Definition}
\theoremstyle{observation}
\newtheorem{observation}{Observation}
\theoremstyle{lemma}
\newtheorem{lemma}{Lemma}
\theoremstyle{claim}
\newtheorem{claim}{Claim}
\theoremstyle{theorem}
\newtheorem{theorem}{Theorem}
\theoremstyle{corollary}

\theoremstyle{remark}

\makeatletter
\renewenvironment{proof}[1][\proofname] {\par\pushQED{\qed}\normalfont\topsep6\p@\@plus6\p@\relax\trivlist\item[\hskip\labelsep\bfseries#1\@addpunct{.}]\ignorespaces}{\popQED\endtrivlist\@endpefalse}
\makeatother
\usepackage{caption}
\makeatletter
\lst@InstallKeywords k{attributes}{attributestyle}\slshape{attributestyle}{}ld
\makeatother

\lstdefinestyle{mystyle}{
  commentstyle=\color{codegreen},
  keywordstyle=\color{magenta},
  numberstyle=\tiny\color{codegray},
  stringstyle=\color{codepurple},
  basicstyle=\footnotesize,
  breakatwhitespace=false,   
  frame=single,
  breaklines=true,                 
  captionpos=b,                    
  keepspaces=true,                 
  numbers=left,                    
  numbersep=5pt,                  
  showspaces=false,                
  showstringspaces=false,
  showtabs=false,
  tabsize=1,
  moreattributes={fromR,poll,peek,predicate,add,join,getNextReady,getStatesIndexes,get,process,initialize,hash},
  attributestyle = \bfseries\color{Blue},
  literate={\ \ }{{\ }}1
}

\usepackage[normalem]{ulem}
\usepackage{amsmath}
\usepackage{float}

\newcommand{\Window}{\ensuremath{\Gamma}}
\newcommand{\window}{\ensuremath{\gamma}}
\newcommand{\WA}{\ensuremath{\textit{WA}}}
\newcommand{\WS}{\ensuremath{\textit{WS}}}
\newcommand{\keybysingle}{\ensuremath{f_K}}
\newcommand{\keybysingleinmath}{\keybysingle}

\newcommand{\aggop}{$A$}
\newcommand{\joinop}{$J$}
\newcommand{\watermarkof}[1]{W_{#1}^\omega}
\newcommand{\watermarkofinmath}[1]{$\watermarkof{#1}$}

\usepackage[ruled,vlined,linesnumbered,norelsize,noend]{algorithm2e}
\usepackage{soul}
\usepackage{hyperref}

\usepackage{array}
\usepackage{bm}

\renewenvironment{description}[1][0pt]
  {\list{}{\labelwidth=0pt \leftmargin=#1
   }}
  {\endlist}
  
\usepackage{threeparttable}
\usepackage{mathtools}
\usepackage{tabularx}

\usepackage{color, colortbl}
\definecolor{Gray}{gray}{0.9}

\usepackage[subtle]{savetrees}

\SetCommentSty{mycommfont}
\let\oldnl\nl
\newcommand{\nonl}{\renewcommand{\nl}{\let\nl\oldnl}}

\DeclarePairedDelimiter\floor{\lfloor}{\rfloor}

\newcommand{\typeof}[1]{T(#1)}

\newcommand{\fusingLogicalAssumption}{\textbf{P1}}
\newcommand{\feedToManyAssumption}{\textbf{P2}}
\newcommand{\cyclicGraphsAssumption}{\textbf{P3}}

\newcommand{\watermarkDAssumption}{\textbf{C1}}
\newcommand{\UWMTemporaryAssumption}{\textbf{C2}}
\newcommand{\DWMTemporaryAssumption}{\textbf{C3}}

\newcommand{\Encapsulate}{Embed}
\newcommand{\EncapsulateAbbr}{\ensuremath{E}}
\newcommand{\EncapsulateFMAbbr}{\ensuremath{E_{FM}}}
\newcommand{\EncapsulateJAbbr}{\ensuremath{E_{J}}}
\newcommand{\Duplicate}{Unfold}
\newcommand{\DuplicateAbbr}{\ensuremath{X}}

\newcommand{\Dedicated}{Dedicated}
\newcommand{\AggBased}{AggBased}

\newcommand{\coderefoneline}[2]{List.\ref{#1},L\ref{#2}}
\newcommand{\coderefonelineDouble}[3]{List.\ref{#1},L\ref{#2},L\ref{#3}}
\newcommand{\codereftwolines}[3]{List.\ref{#1},L\ref{#2}-\ref{#3}}
\newcommand{\codereftwolinesDouble}[5]{List.\ref{#1},L\ref{#2}-\ref{#3},L\ref{#4}-\ref{#5}}

\DeclareMathSymbol{:}{\mathord}{operators}{"3A}

\usepackage{tikz}
\newcommand\encircle[1]{%
  \tikz[baseline=(X.base)] 
    \node (X) [draw, shape=rectangle, inner sep=1, draw=blue!80, thick] {\strut #1};}

\begin{document}
\title{On the Semantic Overlap of Operators\\
in Stream Processing Engines}

\author{Vincenzo Gulisano}
\affiliation{
 \institution{Chalmers University of Technology}
 \city{Gothenburg}
 \country{Sweden}
}
\email{vincenzo.gulisano@chalmers.se}

\author{Alessandro Margara}
\affiliation{%
  \institution{Politecnico di Milano}
 \city{Milano}
  \country{Italy}}
\email{alessandro.margara@polimi.it}

\author{Marina Papatriantafilou}
\affiliation{
 \institution{Chalmers University of Technology}
 \city{Gothenburg}
 \country{Sweden}
}
\email{ptrianta@chalmers.se}

\date{\today}

\begin{abstract}
Stream processing is extensively used in the IoT-to-Cloud spectrum to distill information from continuous streams of data.
Streaming applications usually run in dedicated Stream Processing Engines (SPEs) that adopt the DataFlow model, which defines such applications as graphs of operators that, step by step, transform data into the desired results. As operators can be deployed and executed independently, the DataFlow model supports parallelism and distribution, thus making streaming applications scalable.

Today, we witness an abundance of SPEs, each with its set of operators. In this context, understanding how operators' semantics overlap within and across SPEs, and thus which SPEs can support a given application, is not trivial.
We tackle this problem by formally showing that common operators of SPEs  can be expressed as compositions of a single, minimalistic Aggregate operator, thus showing any framework able to run compositions of such an operator can run applications defined for state-of-the-art SPEs.
The Aggregate operator only relies on core concepts of the DataFlow model such as data partitioning by key and time-based windows, and can only output up to one value for each window it analyzes.
Together with our formal argumentation, we empirically assess how an SPE that only relies on such an operator
compares with an SPE offering operator-specific implementations, as well as study the performance impact of a more expressive Aggregate operator by relaxing the constraint of outputting up to one value per window.

The existence of such a common denominator not only implies the portability of operators within and across SPEs but also defines a concise set of requirements for other data processing frameworks to support streaming applications.
\end{abstract}

\maketitle


\ifdefempty{\vldbavailabilityurl}{}{
\vspace{.3cm}
\begingroup\small\noindent\raggedright\textbf{Artifact Availability:}\\
The source code, data, and/or other artifacts have been made available at \url{\vldbavailabilityurl}.
\endgroup
}

\section{Introduction}\label{sec:introduction}

Stream processing is used to analyze streams of data in near real-time and plays  a central role in the data processing stack of many companies for  informed decision-making based on fresh data.

Dedicated software components, namely Stream Processing Engines (SPEs), tackle the complexity of defining stream processing tasks and executing them in an efficient and scalable manner.
To do so, virtually all distributed SPEs today build on the DataFlow programming and execution model~\cite{akidau2015dataflow}.  This model defines streaming computations as graphs of operators, where each operator transforms input streams into output streams and feeds them to downstream operators.
This approach simplifies parallelism and distribution, as different operators can be deployed on one or more machines.  Multiple instances of the same operator may also run in parallel, each analyzing a portion of the input stream.

Despite all sharing the same DataFlow foundation, each SPE brings its own (vast) library of operators. Understanding how operators' semantics overlap within and across SPEs, and thus which SPEs can support a given streaming application, is not trivial.
Unraveling the semantic overlap between operators and defining a concise core set of operators, shared by all SPEs, that can be used to enforce other operators' semantics, can bring substantial benefits from both a theoretical and a practical viewpoint:

\begin{enumerate}[leftmargin=*]
\item Within and across SPEs, it opens up for polyglot data pipelines in which individual operators or sets of operators can be deployed in the SPE that best suits the users' needs.
\item Developers may reduce their stream processing tasks to equivalent ones that adopt a minimal set of operators, thus simplifying the validation and reasoning on the correctness of applications.
\item By supporting such a core set of operators, Big Data frameworks that are not originally designed as SPEs, can also support Dataflow-based applications.
\item Developers may use a minimal set of operators as a reference to
  define, document, and validate the semantics of new operators.
\end{enumerate}

With this paper, we show that with a single, minimalistic operator that all SPEs implement, and a few basic concepts of the DataFlow model (namely event-time, watermarks, key-by data partitioning, and basic non-nested loops), all common operators found in state-of-the-art SPEs can be expressed as a composition of multiple instances of such an operator.
Specifically, we only consider an Aggregate operator that analyzes a time-based window and outputs up to one result every time the window slides.

We provide formal proofs for our claims, and we conduct an empirical evaluation to assess the performance of an SPE that only relies on an Aggregate operator and that only includes the aforementioned basic Dataflow constructs.
As part of our contribution, we also show that slightly extending the capabilities of an SPE beyond the minimal set we identified brings a significant benefit in terms of performance.

In summary, this paper contributes to the research on supporting stream processing in complex data pipelines by:

\begin{enumerate}[leftmargin=*]
\item formally showing that all common operators found in state-of-the-art SPEs can be expressed by compositions of multiple instances of a single, minimalistic Aggregate operator, which only relies on basic concepts of the DataFlow model;
\item assessing the performance of a reference SPE that only relies on such minimalistic Aggregate operator;
\item evaluating the performance gains of relaxing some assumptions about the capabilities of the Aggregate forming such a core set.
\end{enumerate}

Organization: \autoref{sec:prel} presents background concepts on stream processing and the DataFlow model; \autoref{sec:smps} defines the research problem and presents the general approach we follow, \autoref{sec:native} proves the equivalence of common stream processing operators to compositions of a minimalistic Aggregate operator; \autoref{sec:relax} discusses relaxation and possible extensions to our work; \autoref{sec:evaluation} studies the performance analysis of an SPE that only adopts an Aggregate operator; \autoref{sec:related}  discusses research work that is related to our study; \autoref{sec:conclusions} provides conclusive remarks.

\section{Preliminaries}
\label{sec:prel}

\subsection{Stream processing basics}\label{sec:dsbasics}

According to DataFlow~\cite{akidau2015dataflow}, a \textit{stream} $S$ is an unbounded sequence of \textit{tuples}.
Each tuple 
is defined as a list of attribute-value pairs $\left<\tau:v_{0},a_{1}:v_{1},\dots,a_{n}:v_{n}\right>$. 
Streams are homogeneous, meaning that every tuple $t$ of the same stream $S$ has the same list of attributes.  We call this list the \textit{type} of $t$ or $S$, and we use $\typeof{t}$ to denote the type associated with $t \in S$, or $\typeof{S}$ to denote the type associated with any tuple in $S$.
We assume that a special timestamp attribute ($\tau$) is always included in the type of a tuple.
We use $t[i]$ to denote $t$'s attribute at index $i$, with indexes starting at 0 (i.e., $t[0]=t.\tau$), while $t[i:j]$ denotes the list of attributes from $i$ to $j$ (inclusive).
Given an attribute-value pairs list $L$, $\left< L \right>$ denotes the tuple carrying such a list. 
Given two attribute-value pairs lists $L_1$ and $L_2$, $L_{1}{}^\frown L_2$ denotes their concatenation.

\textit{Stream processing queries} (or simply queries) are composed of \textit{ingresses}, \textit{operators}, and \textit{egresses}.
Ingresses forward \textit{ingress tuples} (e.g., events reported by sensors or other applications) to operators, the basic units manipulating tuples. 
Operators connected in a directed graph 
process and forward/produce tuples;
eventually, \textit{egress tuples} are fed to egresses, which deliver results to end-users or other applications.
As we further explain in \autoref{ssc:snparallelism}, multiple copies of the same operator can be deployed within the same graph, each analyzing a portion of a given stream, e.g., that composed of tuples sharing the same key in \textit{key-by} parallelism.

As an ingress tuple $t$ corresponds to an event, $t.\tau$ is the \emph{event time} set by the ingress, indicating when the event took place.
Operators set $t_o.\tau$ of each output tuple $t_o$ according to their semantics, as explained later. 
Other attributes are set by user-defined functions.
Event time is expressed in time units from a given epoch, and progresses in SPE-specific discrete $\delta$ increments (e.g., milliseconds~\cite{flink}). 

As mentioned in \autoref{sec:introduction}, we show that a minimalistic operator named Aggregate, commonly provided by SPEs~\cite{flink,storm,liebre}, and a few core concepts of the DataFlow model suffice to enforce the semantics of other common streaming operators: Filter, Map, FlatMap and Join, introduced next for self-containment.

The first differentiation between these operators is on whether their analysis is \textit{stateless} or \textit{stateful}.
FlatMap, Filter, and Map are stateless operators since they do not maintain any state that evolves according to the tuples they process:

\begin{description}
\item[FlatMap] $S_O = FM(S_I,f_{FM})$ is a general-purpose operator that takes as input a single stream $S_I$ and a function $f_{FM}$, and produces a single stream $S_O$.  Specifically, it invokes function $f_{FM}$ on each input tuple $t_i \in S_I$: the function may produce zero, one, or more output tuples, which are appended to the output stream $S_O$. Notice $f_{FM}$ may change the type of input tuples.  Also notice that $f_{FM}$ does not set the $\tau$ in an output tuple $t_o$ since the latter is set by $FM$, so that, for a $t_o$ produced from $t_i \in S_I$, $t_i.\tau = t_o.\tau$.
\item[Filter] $S_O = F(S_I,f_C)$ forwards each tuple $t_i \in S_I$ to $S_O$ if $f_C(t_i)$ holds.
Note that $\typeof{S_I}=\typeof{S_O}$ and that $t_i=t_o$ for a tuple $t_o$ forwarded upon the processing of $t_i$.
\item[Map] $S_O = M(S_I,f_M)$, for each tuple $t_i \in S_I$, forwards in $S_O$ the tuple $t_o$ based on $f_M(t_i)$.
As also done by $FM$, $f_M$ does not set the $\tau$ in an output tuple $t_o$ since the latter is set by $M$, so that, for a $t_o$ produced from $t_i \in S_I$, $t_i.\tau = t_o.\tau$.
\end{description}

Stateful operators produce results from a state, dependent on one or more tuples.
Such a state can be defined in many ways.  In this paper, we mainly target stateful operators defined over delimited groups of tuples called \textit{time-based windows} (or simply windows), which are most commonly provided by SPEs~\cite{flink,storm,liebre}: we consider \textit{Aggregates} and \textit{Joins} over windows. In \autoref{sec:relax}, we elaborate on how our analysis can be extended to more general definitions of state.

For conciseness, we use the notation $\Window(\WA,\WS,S_I,\keybysingle,L)$ to indicate a window specified by the following parameters:
\begin{description}
\item [Window Advance (\WA{}), Size (\WS{})] define the epochs $[\ell \WA, \ell \WA+ \WS)$, with $\ell \in \mathbb{N}$, covered by $\Window$.
We refer to one such epoch as window \textit{instance} \window.
If $\WA<\WS$, consecutive window instances overlap, $\Window$ is called \textit{sliding}, and a tuple can fall into several window instances. If $\WA=\WS$,  $\Window$ is called \textit{tumbling} and each tuple falls in exactly one window instance. 
We write $t \in \window$ to denote that $t$ falls in $\window$. 
\item [Input stream $\bm{S_I}$] is the input stream fed to $\Window$.
\item [Key-by attribute $\bm{\keybysingleinmath{}}$] specifies which subset (possibly empty) of attributes from $\typeof{S_{I}}$ to use to maintain distinct window instances for tuples that share the same \textit{key}. Note that $\keybysingleinmath{}$ affects the way in which the operator maintaining $\Window$ can be parallelized (see~\autoref{ssc:snparallelism}). 
\item [Allowed Lateness $\bm{L}$] which is used to decide whether a tuple $t$ falling in $\window$ but received by the operator maintaining $\Window$ after such operator has produced a result for $\window$ should still be added to $\window$, potentially resulting in a new (or updated) output tuple\footnote{We discuss $L$ in \autoref{sec:latearrivals}, after covering correctness conditions in \autoref{ssc:correctness}.}.
\end{description}

We refer to the set of tuples falling in a window instance $\window$ as $\window.\zeta$ and refer to individual tuples in $\window.\zeta$ as if $\window.\zeta$ is maintained as a list (i.e., $\window.\zeta[0]$ is the tuple at index 0). We refer to the event time of $\window$'s left boundary (inclusive) as $\window.l$. The right boundary of $\window$ (exclusive) is computed as $\window.l+\WS$.
As common in related works~\cite{flink,storm,beam}, when an output tuple $t_{o}$ is created in connection to a window instance $\window$, $t_{o}.\tau$ is set to $\window.l+\WS-\delta$. Since $\window$'s right boundary is exclusive:
\begin{observation}
\label{obs:outtuplestimestamp}
For any output tuple $t_{o}$ produced from a window instance in which input tuple $t_{i}$ falls, $t_{o}.\tau \geq t_{i}.\tau$.
\end{observation}

\noindent We consider the following stateful operators:
\begin{description}
    \item[Aggregate] $S_O{=}A(\Window(\WA,\WS,S_I,\keybysingle,L),f_O)$ defines function $f_O(\window)$ to compute the values of an output $t_o$ from $\window$ (except $\tau$) and forwards $t_o$ if $f_O$ does not return $\emptyset$. Note that 
    $t_o.\tau$, as aforementioned, is set by $A$ depending on the window instance $\window$ on which $f_O$ is run. 
    \item[Join] $S_O{=}J(\Window(\WA,\WS,S_{I_1},\keybysingle^1,L),\Window(\WA,\WS,S_{I_2},\keybysingle^2,L)$, $f_P)$ defines $f_P$ to match 
    pair of tuples $t_1 \in S_{I_1}$ and $t_2 \in S_{I_2}$ that fall in aligned windows $\window_1$ and $\window_2$ according to $\WA$ and $\WS$, so that $\keybysingle^1(t_1)=\keybysingle^2(t_2)$. If $f_P(t_1,t_2)$ holds, $J$ forwards $\left< \window_1.l+\WS-\delta ^\frown t_1{}^\frown t_2 \right>$.
\end{description}

In the remainder, we use the following definition:

\begin{definition}
\label{def:successor}
We say tuple $t'$ is a \textit{successor} $t$, or that $t'$ belongs to the finite set of successors of $t$, and write $t' \in succ(t)$ if $t'$ is output by an operator $O$ upon processing of $t$ or a successor of $t$. We also use the notation $succ(T)$ when $T$ is a set of tuples processed by $O$.  
\end{definition}

\subsection{Shared-nothing parallelism}
\label{ssc:snparallelism}

As mentioned in \autoref{sec:dsbasics}, SPEs parallelize the execution of an operator by deploying multiple copies of such an operator, maximizing processing throughput and/or minimizing processing latency. 

\begin{figure}[ht!]
\includegraphics[width=\linewidth]{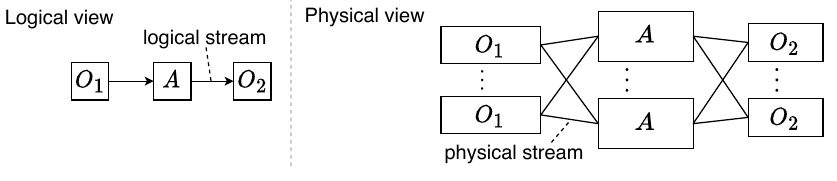}
\caption{Logical/physical views of an $A$ operator and its upstream ($O_1$) and downstream ($O_2$) peers.}\label{fig:sn}
\end{figure}

To achieve this, SPEs let users define operators as \textit{logical}, and later convert them into \textit{physical instances} (\autoref{fig:sn}).
Since stateless operators process tuples one by one, the data fed to multiple instances of the same logical operator can be shuffled or fed in a round-robin fashion. 
For logical stateful operators, though, SPEs run one or more physical instances leveraging key-by parallelism, splitting the data sent to such instances so that tuples sharing the same $\keybysingle$ value (for $A$) or $\keybysingle^1,\keybysingle^2$ values (for $J$) are correctly processed by the same instance (see Section~\ref{ssc:correctness} for more on correctness).
Each logical stream connecting a pair of logical operators (or an ingress/egress) is converted into one or more physical streams.

\subsection{Correctness conditions}
\label{ssc:correctness}

When deploying and running $F$, $M$, $FM$, $A$, and $J$ operators, users expect SPEs to enforce such operators' semantics correctly.
Since $F$, $M$, and $FM$ process tuples one by one without an evolving state, correct semantics are trivially enforced, as long as each tuple is processed exactly once.
$A$ and $J$ require greater care to correctly enforce their semantics, though.
Leaving aside the existence of late arrivals (see \autoref{sec:latearrivals}), their correct execution can be defined as follows.
\begin{definition}
\label{def:correctness}
$A/J$ execution is correct if, according to $A/J$ specifications, any subset of $A/J$'s input tuples from $A/J$'s input stream(s) that could jointly contribute to an output tuple is indeed processed together and results in such an output tuple (if any). 

\end{definition}

For \aggop{}, Definition~\ref{def:correctness} implies that all the input tuples falling into window instance $\window$ should be jointly processed by $f_O$.
For \joinop{}, it implies that any pair of tuples $t_{1}\in \window_1$ and $t_{2} \in \window_2$ is processed by $f_P$ if $\window_1.l=\window_2.l$ and $\keybysingle^1(t_{1})=\keybysingle^2(t_{2})$.
As discussed in~\cite{gulisano2020role} correct execution for $A/J$ builds on consistently maintaining their \textit{watermarks}:
\begin{definition}
\label{def:watermark}
The watermark $\watermarkof{A}/\watermarkof{J}$ of $A/J$ at wall-clock time\footnote{From here on, we only differentiate wall-clock time (or simply time) from event time if such a distinction is not clear from the context.} $\omega$ is the earliest event time a tuple $t_i$ fed to $A/J$ can have from time $\omega$ on (i.e., $t^{i}.\tau \ge \watermarkof{A}/\watermarkof{J}, \forall t_{i}$ processed from $\omega$ on).
\end{definition}

In the literature~\cite{flink,gulisano2020role}, watermarks are commonly maintained assuming ingresses periodically output watermarks as special tuples.
They serve as notifications of how event-time advances from the perspective of ingresses: operators use them as barriers that allow to (1)~make progress even in absence of explicit tuples, (2)~reorder tuples received from out-of-timestamp-order streams.
Upon receiving a watermark, $A/J$ stores the watermark's time, updates $\watermarkof{A}/\watermarkof{J}$ to the smallest value among those in the set comprised of the latest watermark from each input stream, and propagates $\watermarkof{A}/\watermarkof{J}$.

Upon reception of a watermark that increases \watermarkofinmath{A}, $A$ can output the results of all window instances whose right boundary is not greater than \watermarkofinmath{A} (i.e., invoke $f_O$ on any $\window | \window.l+\WS\leq \watermarkof{A}$) since no more tuples will fall in such windows.
Given Observation~\ref{obs:outtuplestimestamp}, this implies that $A$ can also forward $\watermarkof{A}$ to its downstream peers once such results are output.
Likewise, a $J$ operator can use watermark \watermarkofinmath{J} to safely discard window instances that cannot produce any further result.  Specifically,  a $J$ operator 
can safely discard any pair of window instances $\window_1, \window_2$ defined over $S_{I_1}$ and $S_{I_2}$, respectively, for which it holds that $\window_1.l=\window_2.l$ and $\window_1.l+\WS \leq \watermarkof{J}$.
Given Observation~\ref{obs:outtuplestimestamp}, \watermarkofinmath{J} can be forwarded upon each update of its value.

\subsection{Handling late arrivals}
\label{sec:latearrivals}

As introduced in \autoref{sec:dsbasics}, $\Window$ defines $L$ to handle late arrivals. 
Tuple $t$ is a late arrival for operator $O$ if $t.\tau < W^\omega_O$ when, at time $\omega$, $O$ processes $t$. 
According to the Dataflow model, 
$t$ is processed, added to $\window$, and can result in an output tuple (potentially representing an update of a previous output tuple) if $\window.l + \WS \leq W_O^\omega + L$ at $\omega$.
In this case, the purging of a window instance $\window$ is delayed by $L$, accounting for potential late arrivals.
Note that, by setting $L=0$, $O$ will not process any late arrivals.
If $L>0$ and watermarks are forwarded by $O$ as described in \autoref{ssc:correctness}, though, results produced by $O$ could be late arrivals for $O$'s downstream peers. 
For compact notation, we omit $L$ for a $\Window$ if $L=0$. In the remainder, we also make use of the following definition:

\begin{definition}
\label{def:triggered}
We say tuple $t$ is an output tuple \textit{triggered} upon the growth of $O$'s watermark to $W_O$ and write $t \in trig(W_O)$ if $t$ is produced and forwarded by $O$ when $O$'s watermark grows to $W_O$.
\end{definition}

\section{Problem Definition and Approach}\label{sec:smps}

This work has two main goals.
First, to show that the semantics of $F$, $M$, $FM$, and $J$ operators can be enforced by composing $A$ operators (see \autoref{sec:dsbasics}).
Second, to compare the performance of \textit{\Dedicated{}} implementations of $F$, $M$, $FM$, and $J$ (i.e., the implementations offered by an SPE), and \textit{\AggBased{}} implementations based on a composition of $A$ operators, based on common metrics like throughput and latency~\cite{cederman2013concurrent,scalejoin,teubner2011soccer,gulisano2022stretch}, as further discussed in \autoref{sec:evaluation}.

While focusing mainly on showing that the basic $A$ operator described in \autoref{sec:dsbasics} suffices for enforcing $F$, $M$, $FM$, and $J$ operators, we also consider in \autoref{sec:aplus} a variation of $A$ that is semantically richer, and show how it can simplify the definition of such $F$, $M$, $FM$, and $J$. Also, while focusing mainly on $F$, $M$, $FM$, and $J$ operators with time-based windows, we show in \autoref{sec:otherstate} it is possible for compositions of $A$ operators to maintain arbitrary states for other streaming operators too, something we intend to explore in future work.

In our work, we consider SPEs for which the following holds:

\begin{description}
    \item[\fusingLogicalAssumption] Physical streams sharing the same type can be fed to the same $A$ operator. If such physical streams belong to the logical streams $S_{I_1},S_{I_2},\ldots$ we write $\{S_{I_1},S_{I_2},\ldots\}$ to refer to the logical stream fed to $A$ and $\typeof{S_{I_1}}$ to refer to the shared type of such streams.
    \item[\feedToManyAssumption] Physical/logical streams can feed one or more $A$ operators, delivering the same tuples/watermarks in the same order.
    \item[\cyclicGraphsAssumption] An $A$ operator can define a loop to iterate over its own output stream. A watermark forwarded by $A$ is not fed again to $A$ through such a loop (since $A$ itself produced it).
\end{description}

About \fusingLogicalAssumption, note that SPEs like Apache Flink~\cite{flink} automatically merge physical streams from parallel instances of the same operator (see~\autoref{ssc:snparallelism}), but require a call to the \texttt{union} method for physical streams from different logical streams. Since such a call can be handled automatically at query compile time, we do not differentiate the two cases.

About \cyclicGraphsAssumption, note that an output $t_o$ from $A$ fed to $A$ via a loop is always a late arrival since it holds that $\watermarkof{A}>t_o.\tau$ for $t_o$ to be output (see \autoref{ssc:correctness}) and that, upon processing of such $t_o$, any resulting output will also constitute a late arrival for $A$'s downstream peers. We refer to Appendix~\ref{app:A} for an extended discussion and \cite{trofimov2022substream} for further details on the complexities of loops in streaming queries.

We limit our scope to the handling of late arrivals for $A$s fed via a loop, thus assuming an \AggBased{} $J$ does not need to handle late arrivals (i.e., $L=0$ for a $J$), and assuming the next conditions hold:
\begin{description}
\item[\watermarkDAssumption] each stream $S$ delivers watermarks periodically, with a maximum event-time distance $D$ between consecutive watermarks. If the first tuple $t^0 \in S$ precedes the first watermark $W^0$, then $W^0 - t^0.\tau \leq D$,
\item[\UWMTemporaryAssumption] the SPE running $A$ does not prevent any $t$ fed to $A$ through a loop from being processed on the basis of being a late arrival, and
\item[\DWMTemporaryAssumption] the SPE running $A$ forwards all the outputs and watermarks of $A$ to $A$'s downstream peers so that no output is a late arrival. 
\end{description}
 
About $\watermarkDAssumption{}$ note that if an $A$ operator is fed an ingress stream for which $\watermarkDAssumption{}$ holds, a distance $D$ exists for $A$'s output too. By extension, if an \AggBased{} $F$, $M$, $FM$ or $J$ fed $S$ produces a stream for which a  $D$ can be defined according to $\watermarkDAssumption{}$, $\watermarkDAssumption{}$ extends to such $F$, $M$, $FM$ or $J$ and downstream $F$, $M$, $FM$ or $J$ peers too.  
Also, note \UWMTemporaryAssumption{} and \DWMTemporaryAssumption{} are given here in a simplified form for ease of exposition. We refer to \autoref{sec:loops} for a more detailed formulation based on $A$'s watermarks. In \autoref{sec:instrumentedstreamsimplementation}, we provide algorithms to enforce \UWMTemporaryAssumption{} and \DWMTemporaryAssumption{} in case the SPE in use does not support them natively, while in \autoref{sec:aplus} we discuss relaxations to our model that do not require \cyclicGraphsAssumption{}, \watermarkDAssumption{}, \UWMTemporaryAssumption{}, and \DWMTemporaryAssumption{}.

For ease of notation, and without lack of generality, we center our discussion on logical operators and streams (see \autoref{ssc:snparallelism}).
Note the parallel instances of the same logical operator are fed the same watermarks and can thus run their analysis, produce results, and forward watermarks as we describe in the remainder.
Finally, for ease of exposition, all lemmas' proofs are found in Appendix~\ref{app:proofs}.

\renewcommand*{\algorithmcfname}{Listing}

\section{Enforcing Common Operators' Semantics with $A$ operators}
\label{sec:native}

In this section, we show that the semantics of $F$, $M$, $FM$, and $J$ operators can be enforced by composing $A$ operators.
We begin by noting that the semantics of $FM$ operators, able to produce an arbitrary number of tuples upon processing of an input tuple $t$, encapsulates those of $F$ operators, which produce 0 or 1 tuples upon the processing of $t$, and those of $M$ operators, which produce exactly 1 tuple upon the processing of $t$.
Hence, showing that the semantics of an $FM$ can be enforced by composing $A$ operators shows $A$'s semantics are sufficient to enforce those of $F$ and $M$ too.
Regarding $J$ and $FM$, an intrinsic challenge, considering the minimal $A$ that produces 0 or 1 tuples per window instance, is that $J$ and $FM$ may produce multiple tuples from a single tuple or window instance $\window$, respectively.
As we show, the \emph{key idea} to address this is to decompose this functionality into encapsulating the content of multiple tuples within one tuple and later unfolding such content into individual tuples.
To do so, we introduce two \emph{auxiliary operators: \Encapsulate{} (\EncapsulateAbbr{}) and \Duplicate{} (\DuplicateAbbr{})}, to embed several tuples into one and to unfold and output them one by one, respectively. \EncapsulateAbbr{} and \DuplicateAbbr{} exploit basic building blocks of the DataFlow model, namely data partitioning and loops.

Elaborating on these ideas, this section first overviews $\EncapsulateAbbr{}$ and $\DuplicateAbbr{}$ and discusses how they support $FM$ and $J$ semantics in \autoref{sec:eandd}; it subsequently 
shows that the semantics of the $\EncapsulateAbbr{}$ encapsulating an $FM$ can be enforced with an $A$ in \autoref{sec:eFM},
while the semantics of the $\EncapsulateAbbr{}$ encapsulating a $J$ can be enforced composing $A$s in \autoref{sec:eJ},
as well as that the semantics of the $\DuplicateAbbr{}$ operator can be enforced composing $A$s in \autoref{sec:loops}, and
proposes algorithms to meet \UWMTemporaryAssumption{} and \DWMTemporaryAssumption{} for SPEs not supporting them natively.

\subsection{The \Encapsulate{} ($\EncapsulateAbbr{}$) and \Duplicate{} ($\DuplicateAbbr{}$) Operators}
\label{sec:eandd}

\begin{figure}[h]
\includegraphics[width=\linewidth]{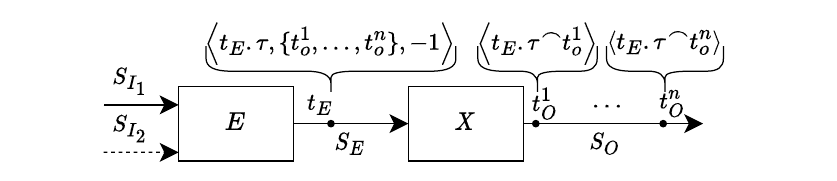}
\caption{Composition of \Encapsulate{} (\EncapsulateAbbr{}) and \Duplicate{} (\DuplicateAbbr{}) that can enforce $FM$ and $J$ semantics.}
\label{fig:eandd}
\end{figure}

Figure~\ref{fig:eandd} overviews how we compose \EncapsulateAbbr{} and \DuplicateAbbr{} operators to enforce the semantics of $FM$/$J$ operators. For $FM$, \EncapsulateAbbr{} output a single tuple $t_E$ embedding all the results that derive from processing $t_i$, and value $-1$ a special value used for termination, upon processing an input tuple $t_i$. For $J$, \EncapsulateAbbr{} embeds in $t_E$ all the matching pairs from two aligned window instances (see \autoref{sec:dsbasics}). The subsequent \DuplicateAbbr{} operator unwraps $t_E$ and outputs all tuples $\{t^1_o,\ldots,t^n_o\}$ that were included in $t_E$.  Internally, it uses a loop that iteratively extracts tuples from $t_{E}$ until reaching the terminator value $-1$.
More formally:

\begin{itemize}[leftmargin=*]
    \item \EncapsulateAbbr{} takes one input stream $S_{I_1}$ and an optional input stream $S_{I_2}$, possibly with a different type\footnote{$S_{I_2}$ is optional since $FM$ has a single logical input stream (\autoref{sec:dsbasics}).} (i.e., $\typeof{S_{I_1}}$ and $\typeof{S_{I_2}}$ can differ).
    \item \EncapsulateAbbr{} produces one output stream $S_{\EncapsulateAbbr{}}$. A tuple $t_{\EncapsulateAbbr} \in S_{\EncapsulateAbbr{}}$ carries its timestamp $t_{\EncapsulateAbbr}.\tau$, a set of tuples $\{t^1_o,\ldots,t^n_o\}$, and value $-1$. As we explain in \autoref{sec:loops}, value $-1$ identifies $t_{\EncapsulateAbbr}$ as output of \EncapsulateAbbr{}.
    \item \DuplicateAbbr{} produces one output stream $S_O$. Upon reception of tuple $t_{\EncapsulateAbbr}$, it outputs $\left< t_{\EncapsulateAbbr}.\tau ^\frown t^1_o \right>$, $\ldots$, $\left< t_{\EncapsulateAbbr}.\tau ^\frown t^n_o \right>$.
\end{itemize}

For an $\EncapsulateAbbr{}$ and $\DuplicateAbbr{}$ pair (see \autoref{fig:eandd}) to enforce the semantics of an $FM$, $\EncapsulateAbbr{}$ should encapsulate $f_{FM}$ and output the tuples produced by $f_{FM}(t)$ in a $t_{\EncapsulateAbbr}$ so that $t_{\EncapsulateAbbr}.\tau=t.\tau$, for \DuplicateAbbr{} to later unwrap them. More concretely:

\begin{claim}
\label{obs:eflatmap}
\EncapsulateAbbr{} and \DuplicateAbbr{} enforce the semantics of 
$S_O = FM(S_{I_1},f_{FM})$ 
if 
for each $t \in S_{I_1}$ fed to $E$, there is a $t_{\EncapsulateAbbr}$ such that $t_{\EncapsulateAbbr}.\tau=t.\tau$ and $f_{FM}(t)$ is carried in $t_{\EncapsulateAbbr}$'s second attribute (i.e.,  $t_{\EncapsulateAbbr}[1]$, see \autoref{sec:dsbasics}).
\end{claim}

For $\EncapsulateAbbr{}$ and $\DuplicateAbbr{}$ (see \autoref{fig:eandd}) to enforce $J$'s semantics, $\EncapsulateAbbr{}$ should encapsulate $J$'s parameters and embed all matching tuples from a pair of aligned windows $\window_1$ and $\window_2$ in a $t_{\EncapsulateAbbr{}}$ with $t_{\EncapsulateAbbr{}}.\tau=\window_1.l+WS-\delta$, for such tuples to be later unwrapped by \DuplicateAbbr{}. More concretely:

\begin{sloppy}
\begin{claim}
\label{obs:ejoin}
\EncapsulateAbbr{} and \DuplicateAbbr{} 
enforce the semantics of 
$S_O{=}J(\Window(\WA,\WS,S_{I_1},\keybysingle^1), \Window(\WA,\WS,S_{I_2}, \keybysingle^2), f_P)$ 
if,
for each pair of tuples $t_1 \in S_{I_1}$ and $t_2 \in S_{I_2}$ such that 
$\keybysingle^1(t_1)=\keybysingle^2(t_2)$ and $f_P(t_1,t_2)$, 
and for each pair of windows $\window_1$ and $\window_2$ such that $t_1 \in \window_1$, $t_2 \in \window_2$, and $\window_1.l=\window_2.l$,
\EncapsulateAbbr{} produces an output tuple $t_{\EncapsulateAbbr{}}=\left< \window_1.l+WS-\delta ^\frown \{t^1_o,\ldots,t^n_o\} ^\frown -1 \right>$ 
carrying $t_1{}^\frown t_2$ in $t_{\EncapsulateAbbr{}}[1]$.
\end{claim}
\end{sloppy}

In the remainder, we write $\EncapsulateFMAbbr{}(S_{I_1},f_{FM})$ to refer to the \EncapsulateAbbr{} encapsulating $FM$'s semantics, $\EncapsulateJAbbr{}(\WA, \WS, S_{I_1}, S_{I_2}, \keybysingle^1, \keybysingle^2, f_P)$ for the $\EncapsulateAbbr{}$ encapsulating $J$'s semantics, and $\DuplicateAbbr(S_E)$ for $\DuplicateAbbr{}$.

\subsection{Using an $A$ to Enforce $\EncapsulateAbbr_{FM}$'s Semantics}
\label{sec:eFM}

\begin{theorem}
\label{thm:map}
The semantics of $S_E = \EncapsulateAbbr_{FM}(S_{I_1},f_{FM})$ described in Claim~\ref{obs:eflatmap} can be enforced with the $A$ operator in Listing~\ref{alg:eflatmap}.

\begin{algorithm}[h]
\footnotesize
\SetAlgoLined
\DontPrintSemicolon
\SetKwProg{Proc}{Function}{}{}
\BlankLine
\nonl $S_E = \EncapsulateAbbr{}_{FM}(S_I,f_{FM}) = A(\Window(\delta,\delta,S_I,\typeof{S_I}),f_O)$, where:\;  
\BlankLine
\Proc{$f_O(\window)$}
{
    $T \xleftarrow{} \{\}$ \tcp{Create empty set $T$}
    \For(\tcp*[h]{For each $t' \in \window$}){$t' \in \window.\zeta$}{
        $T \xleftarrow{} T \cup f_{FM}(t')$ \tcp{Add $f_{FM}(t')$ to $T$}
    }
    \If(\tcp*[h]{One or more tuples from $f_{FM}$}){$T\neq\{\}$}{
        \Return $T ^\frown -1$\;
    }\Else(\tcp*[h]{No tuple from $f_{FM}$}){
        \Return $\{\}$\;
    }
}
\caption{\small $A$ operator implementing $\EncapsulateAbbr_{FM}$'s semantics}
\label{alg:eflatmap}
\end{algorithm}
\end{theorem}

Before proving Theorem~\ref{thm:map}, we introduce the following lemma.

\begin{lemma}
\label{obs:tumbling}
If $A$ defines a tumbling $\Window$ with $\WA=\delta$ and $\WS=\delta$, then any $t$ falls in exactly one window instance $\window|\window.l=t.\tau$. Moreover, if $t_i \in \window$ and $t_o$ is produced upon invocation of $f_O(\window)$, then $t_o.\tau=t_i.\tau$.
\end{lemma}

\noindent
Given Lemma~\ref{obs:tumbling} we can now prove Theorem~\ref{thm:map}.

\begin{proof}
(Theorem~\ref{thm:map}) Each tuple $t$ falls in exactly one window instance because $\Window$ is a tumbling window.
If a window instance $\window$ contains more than a tuple, then all the tuples in $\window$ are identical, because $A$ uses the entire list of attributes $T(S_{I_1})$ as key-by.
Since all tuples in $\window$ are identical, each tuple $t' \in \window$ results in the same set of tuples once fed to $f_{FM}$, and $T$ contains the tuples of all such sets, or is empty if $f_{FM}$ does not return any tuple. 
If an output $t_o$ is produced, then $t_o.\tau$ for a \window{} is the same as those contained in $\window.\zeta$ (Lemma~\ref{obs:tumbling}).
Hence, $A$ enforces the semantics of $\EncapsulateAbbr_{FM}$.
\end{proof}

\subsection{Using $A$s to enforce $\EncapsulateAbbr_J$'s Semantics}
\label{sec:eJ}

\begin{theorem}
\label{thm:join}
The semantics of 
$S_E = \EncapsulateAbbr_{J}(\WA,\WS,S_{I_1},S_{I_2},\keybysingle^1,\keybysingle^2,f_P)$
can be enforced by composing $A$s as shown in Figure~\ref{fig:join} and Listing~\ref{alg:join}.

\begin{figure}[h]
\includegraphics[width=\linewidth]{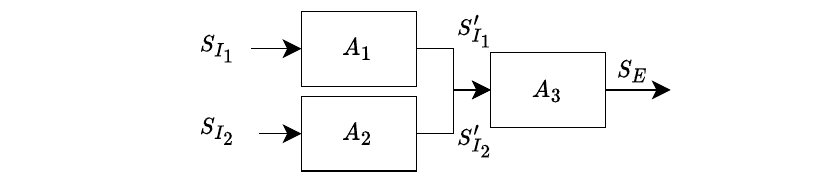}
\caption{Graph of $A$ operators implementing $J$'s semantics.}
\label{fig:join}
\end{figure}

\begin{algorithm}[h]
\footnotesize
\SetAlgoLined
\DontPrintSemicolon
\SetKwProg{Class}{Class}{}{}
\SetKwProg{Proc}{Function}{}{}
\BlankLine
\nonl  $S'_{I_1} = A(\Window(\delta,\delta,S_{I_1},\typeof{S_{I_1}}),f_O)$, where: \tcp{\encircle{$A_1$}}
\BlankLine
\Proc{$f_O(\window)$}
{\label{alg:join:a1start}
    $T \xleftarrow{} \{\}$ \tcp{Create empty set $T$}
    \For(\tcp*[h]{For each $t' \in \window$}){$t' \in \window.\zeta$}{
        $T \xleftarrow{} T \cup t'$ \tcp{Add $t'$ to $T$}
    }
    \Return $T ^\frown \{\}$
}
\nonl\hrulefill\\
\nonl $S'_{I_2} = A(\Window(\delta,\delta,S_{I_2},\typeof{S_{I_2}}),f_O)$, where: \tcp{\encircle{$A_2$}}
\BlankLine
\Proc{$f_O(\window)$}
{
    $T \xleftarrow{} \{\}$ \tcp{Create empty set $T$}
    \For(\tcp*[h]{For each $t' \in \window$}){$t' \in \window.\zeta$}{
        $T \xleftarrow{} T \cup t'$ \tcp{Add $t'$ to $T$}
    }
    \Return $\{\} ^\frown T$\label{alg:join:a2end}
}
\nonl\hrulefill\\
\nonl $S_E = A(\Window(\WA,\WS,\{S'_{I_1},S'_{I_2}\},\keybysingle'),f_O)$, where: \tcp{\encircle{$A_3$}}
\BlankLine
\Proc{$\keybysingle'(t)$}
{\label{alg:join:fk:start}
    \If(\tcp*[h]{$t$ is from $S_{I_1}$}){$t[2]=\{\}$} {
        \Return $\keybysingle^1(T[1][0])$ \tcp{$\keybysingle^1$ of first $t$ in $T[1]$}
    } \Else(\tcp*[h]{$t$ is from $S_{I_2}$}) {
        \Return $\keybysingle^2(T[2][0])$ \tcp{$\keybysingle^2$ of first $t$ in $T[2]$}\label{alg:join:fk:end}
    }
}
\BlankLine
\Proc{$f_O(\window)$}
{\label{alg:join:f_o_start}
    $win1 \xleftarrow{} \{\}$ \tcp{Create list for tuples from $S_{I_1}$}
    $win2 \xleftarrow{} \{\}$ \tcp{Create list for tuples from $S_{I_2}$}
    $T \xleftarrow{} \{\}$ \tcp{Create list for output tuples}
    \For{$t \in \window.\zeta$}{\label{alg:join:iterationstart}
        \If(\tcp*[h]{$t$ is from $S_{I_1}$}){$t[2]=\{\}$} {
            \For(\tcp*[h]{For all $t'$ in $t$}){$t' \in t[1]$}{\label{alg:join:matchings1start}
                \For(\tcp*[h]{For all $t''$ in $win2$}){$t'' \in win2$}{
                    \If(\tcp*[h]{Join $t'$ and $t''$}){$f_P(t',t'')$} {
                        $T \xleftarrow{} T \cup t'^\frown t''$\;\label{alg:join:matchings1end}
                    }
                }
                $win1 \xleftarrow{} win1 \cup t'$ \tcp{Store $t'$}\label{alg:join:stores1}
            }
        } \Else(\tcp*[h]{$t$ is from $S_{I_2}$}) {
            \For(\tcp*[h]{For all $t''$ in $t$}){$t'' \in t[2]$}{\label{alg:join:matchings2start}
                \For(\tcp*[h]{For all $t'$ in $win1$}){$t' \in win1$}{
                    \If(\tcp*[h]{Join $t'$ and $t''$}){$f_P(t',t'')$} {
                        $T \xleftarrow{} T \cup t'^\frown t''$\;\label{alg:join:matchings2end}
                    }
                }
                $win2 \xleftarrow{} win2 \cup t''$ \tcp{Store $t''$}\label{alg:join:iterationend}
            }
        }
    }
    \If(\tcp*[h]{There are results to return}){$T \neq \{\}$} {\label{alg:join:forwardTstart}
            \Return $T ^\frown -1$\;
    } \Else(\tcp*[h]{There are no results to return}) {
        \Return $\{\}$\;\label{alg:join:f_o_end}
    }
}
\caption{\small $A$ operators implementing $\EncapsulateAbbr{}_{J}$'s semantics .}
\label{alg:join}
\end{algorithm}
\end{theorem}
In order to ease the understanding and proof of Theorem~\ref{thm:join}, we can observe that, in Listing~\ref{alg:join}, $A_1$ and $A_2$ output streams sharing the same type, composed of timestamp $\tau$ and two sets.
For $A_1$, the first set carries tuples from $S_{I_1}$ and the second set is empty. Symmetrically, for $A_2$, the first set is empty and the second set carries tuples from $S_{I_2}$ (\codereftwolines{alg:join}{alg:join:a1start}{alg:join:a2end}).
According to \fusingLogicalAssumption{} (see \autoref{sec:smps}) they can thus be both transparently fed to $A_3$.
$A_3$, who carries out the matching of tuples from $S_{I_1}$ and $S_{I_2}$, defines a $\keybysingle'$ function that runs either $\keybysingle^1$ or $\keybysingle^2$ on $t$ depending on whether $t$ comes from $S_{I_1}$ or $S_{I_2}$ (\codereftwolines{alg:join}{alg:join:fk:start}{alg:join:fk:end}), that is, it computes the correct key for the input tuple based on the original stream the tuple comes from.
Since $A_1$ uses all input tuples' attributes as key-by, if more than an input tuple is added to the set carried by an output tuple $t_o$, then all the tuples carried by $t_o$ are identical. A similar observation holds for $A_2$. Hence, although $\keybysingle'$ runs $\keybysingle^1$ or $\keybysingle^2$ on the first tuple carried by an input tuple $t$ in $t[1]$ or $t[2]$, $A_3$'s $\keybysingle'$ consistently assigns tuples to the window instances they fall into based on their key-by value.

$A_3$'s $f_O$ iterates through all the tuples sharing the same key (\codereftwolines{alg:join}{alg:join:iterationstart}{alg:join:iterationend}).
For each such tuple $t'$, it runs a Cartesian product matching the tuples from either $S_{I_1}$ or $S_{I_2}$ carried by $t'$ with previously traversed tuples from the other stream (\codereftwolinesDouble{alg:join}{alg:join:matchings1start}{alg:join:matchings1end}{alg:join:matchings2start}{alg:join:matchings2end}) and stores $t'$, for $t'$ to be compared with other tuples in $\window.\zeta$ traversed after $t'$ (\coderefonelineDouble{alg:join}{alg:join:stores1}{alg:join:iterationend}).
All matching pairs of tuples are stored in $T$.
Finally, if $T$ is not empty, $A_3$ forwards $T$ and $-1$ (\codereftwolines{alg:join}{alg:join:forwardTstart}{alg:join:f_o_end}).

\begin{proof}
(Theorem~\ref{thm:join}) By contradiction; if $\EncapsulateAbbr{}_J$ implementation is not correct, then there exists a pair of tuples $t_1$ and $t_2$ falling in windows $\window_1$ and $\window_2$, respectively, so that $f_P(t_1,t_2)$ holds and $\window_1.l=\window_2.l$ but $t_1{}^\frown t_2$ is not added to $T$ by $A_3$'s $f_O$.
According to the function $f_O$ in \codereftwolines{alg:join}{alg:join:f_o_start}{alg:join:f_o_end} both alternatives can only hold if $t_1$ or $t_2$ do not fall in $\window_1$ or $\window_2$, respectively, or if one or both tuples are not fed to the $A_2$ in the first place, which contradicts the initial assumption.
\end{proof}

\subsection{Using $A$s to enforce $\DuplicateAbbr{}$'s Semantics}
\label{sec:loops}

In \autoref{sec:smps}, we introduced  \watermarkDAssumption{}, \UWMTemporaryAssumption{}, and \DWMTemporaryAssumption{} in relation to the possibility for an $A$ to loop over its output tuples (\cyclicGraphsAssumption{}).
Since \DuplicateAbbr{} relies on \cyclicGraphsAssumption{}, we begin this section re-formulating \UWMTemporaryAssumption{} and \DWMTemporaryAssumption{} in a more precise way (we relied on simplified formulations so far for ease of exposition). The reader can refer to \autoref{sec:instrumentedstreamsimplementation} for algorithms that enforce \UWMTemporaryAssumption{} and \DWMTemporaryAssumption{} in case the SPE in use does not support them natively.
\begin{description}
\item[\UWMTemporaryAssumption] $\watermarkof{A}$ is updated only once $\watermarkof{A}$ does not prevent any $t_0$, produced by $A$ and fed to $A$ via a looping stream $S$, from being processed on the basis of being a late arrival. That is, if $\forall t_o \in S$, $t_o$ is still to be processed by $A$ at $\omega$, and being $\window$ at $A$ a window instance such that $t_o \in \window$, then it holds that $\window.l + \WS \leq \watermarkof{A}+L$, and
\item[\DWMTemporaryAssumption] $A$ feeds $\watermarkof{A}$ to downstream peers after $succ(trig(\watermarkof{A}))$.
\end{description}
Given such refined formulations, we can state the following:

\begin{figure}[ht!]
\includegraphics[width=\linewidth]{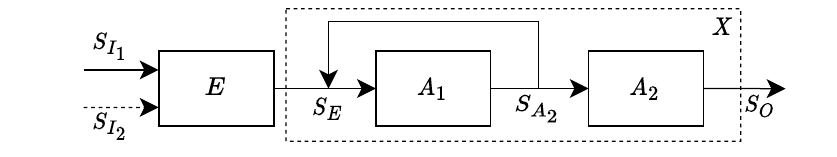}
\caption{Composition of $A$ operators for the \DuplicateAbbr{} operator.}
\label{fig:loops}
\end{figure}

\begin{theorem}
The semantics of a $S_O = \DuplicateAbbr(S_I)$ can be implemented by composing operators $A_1$ and $A_2$ as shown in Figure~\ref{fig:loops} and Listing~\ref{alg:duplicate}, where \watermarkDAssumption{} holds for $S_E$, \UWMTemporaryAssumption{} and \DWMTemporaryAssumption{}  hold for $A_1$, and $L \geq D$.
\label{thm:duplicate}
\end{theorem}

\begin{algorithm}[h]
\footnotesize
\SetAlgoLined
\DontPrintSemicolon
\SetKwProg{Proc}{Function}{}{}
\BlankLine
\nonl $S_{A_2} = A(\Window(\delta,\delta,\{S_{E},S_{A_2}\},\typeof{S_{E}},L),f_O)$, where: \tcp{\encircle{$A_1$}}
\BlankLine
\Proc{$f_O(\window)$}
{
    $t \xleftarrow{} \window.\zeta[0]$\;
    \If(\tcp*[h]{If $t$ from $\EncapsulateAbbr$}){$t[2]=-1$} {\label{alg:duplicate:m2:iffromE}
        $T \xleftarrow{} \{\}$ \tcp{Create empty set}
        \For{$t' \in \window.\zeta$}{
            $T \xleftarrow{} T \cup t'[1]$ \tcp{Fill $T$ with outputs}
        }
        \Return $T^\frown 0$ \tcp{Forward $T$ an index $0$} \label{alg:duplicate:m2:iffromEend}
    } \ElseIf(\tcp*[h]{If $t$ should loop more}){$t[2]<|t[1]|$} {\label{alg:duplicate:m2:checkindex}
        \Return $t[1] ^\frown (t[2]+1)$ \tcp{Increase index}\label{alg:duplicate:m2:increase}
    } \Else(\tcp*[h]{If $t$ done looping}) {
        \Return $\emptyset$\;
    }
}
\nonl\hrulefill\\
\nonl $S_{O} = A(\Window(\delta,\delta,S_{A_2},\typeof{S_{A_2}}),f_O)$, where: \tcp{\encircle{$A_2$}}  
\BlankLine
\Proc{$f_O(\window)$}
{\label{alg:duplicate:a2:f_O_start}
    $t \xleftarrow{} \window.\zeta[0]$\;
    \Return $t[1][t[2]]]$ \tcp{Forward tuple at given index} \label{alg:duplicate:m3:forward}
}
\caption{\small $A$ operators implementing $\DuplicateAbbr{}$'s semantics.}
\label{alg:duplicate}
\end{algorithm}

To ease the understanding and proof of Theorem~\ref{thm:duplicate}, we can observe that, in Listing~\ref{alg:duplicate}, $A_1$ has a tumbling window of size $\delta$ and Allowed Lateness $L$ (see \autoref{ssc:correctness}), and that $S_{A_2}$ feeds both $A_1$ and $A_2$, according to \feedToManyAssumption{}. If an input $t$ comes from $\EncapsulateAbbr$ and thus carries $-1$ as the last value (\coderefoneline{alg:duplicate}{alg:duplicate:m2:iffromE}), $A_1$ forwards a tuple $t_o$ carrying all the output tuples carried by $t$ and value $0$, indicating the next tuple to be produced by $A_2$ is that at index 0. Otherwise, if $t$ comes from $A_1$ itself, $A_1$ forwards a $t_o$ in which the counter at the last attribute is increased by 1 (\coderefoneline{alg:duplicate}{alg:duplicate:m2:increase}).
Note that, if two identical tuples are fed to $A_1$ from $E$, they will end up in the same window instance $\window$, because $A_1$'s $\keybysingle$ selects all tuples' attributes.
Hence, all the $t_o$ tuples carried by such duplicate tuples will be added to the output produced from $\window$. We can also formulate the following lemma when considering how $A_1$ handles duplicate tuples.

\begin{lemma}
\label{thm:noduplicates}
$A_1$ cannot produce duplicates. 
\end{lemma}

For the elements carried by an input tuple $t$ in its second attribute, $A_2$ forwards the one at the index specified by $t$'s third attribute (\coderefoneline{alg:duplicate}{alg:duplicate:m3:forward}).
Since each tuple fed to $A_1$ carries a finite set of $t_o$ tuples, a tuple $t$ output from $A_1$ carries a finite set of $t_o$ tuples too. Hence, the growing index at $t[2]$ will eventually be greater than $|t[1]|$, preventing $t$ from looping for an unbounded number of times through $A_1$ and being fed an unbounded number of times to $A_2$.

\begin{proof}
(Theorem~\ref{thm:duplicate}) 
By contradiction; $t^l=\left< \tau^{l\frown} T ^\frown -1 \right>$ with $T=\{\ldots,t_o^j,\ldots\}$ is fed to $A_1$ but $A_2$ does not produce $t_o^j$.

If $A_2$ does not produce $t_o^j$, then
$A_2$ did not receive $t^*=\left< \tau^{l\frown} T ^\frown j\right>$ \textbf{(1)},
$A_2$ received $t^*$ but as a late arrival that was not processed \textbf{(2)}, or 
$A_2$ received $t^*$ but $t^*$ was not in $\window.\zeta[0]$, since $\window.\zeta[0]$ is the only tuple considered by $A_2$'s $f_O$ (\codereftwolines{alg:duplicate}{alg:duplicate:a2:f_O_start}{alg:duplicate:m3:forward}) \textbf{(3)}.

\textbf{(1)} implies the sequence $\left< \tau ^\frown T ^\frown -1 \right>$, $\ldots$, $\left< \tau ^\frown T ^\frown j-1 \right>$ was not delivered in its entirety to $A_1$. Assuming $\left< \tau ^\frown T ^\frown -1 \right>$ was not delivered contradicts the initial assumption. If $\left< \tau ^\frown T ^\frown -1 \right>$ is processed, because of \UWMTemporaryAssumption{} and \watermarkDAssumption{} and given that all the tuples in the sequence share the same timestamp, \textbf{(1)} results in a contradiction.

\textbf{(2)} implies $t^l$ was fed to $A_1$. Let us refer to the window instance to which $t^l$ falls into at $A_1$ as $\window_l$.
If $t^*$ was produced by $A_1$, then $A_1$ received a watermark $W^m$ such that $W^m\geq t^*.\tau+\delta$ ($\window_l.\tau=t^*.\tau$, see Lemma~\ref{obs:tumbling}). 
Let $W^m$ be the earliest watermark greater than or equal to $t^*.\tau+\delta$, i.e., $W^{m-1}<t^*.\tau+\delta$.
\DWMTemporaryAssumption{} ensures $W^{m-1}$ is the latest watermark fed to $A_2$, because $t^* \in succ(trig(W^m))$.
To be a late arrival for $A_2$, though, $t^*.\tau < W^{m-1}$, which leads to a contradiction.

\textbf{(3)} implies there exist identical tuples fed to $A_2$, because $A_2$ uses all attribute values as key-by. Hence, \textbf{(3)} contradicts Lemma~\ref{thm:noduplicates}.
\end{proof}

\subsection{Handling Watermarks for $\DuplicateAbbr{}$'s $A_1$}
\label{sec:instrumentedstreamsimplementation}

SPEs that offer limited support for cyclic graphs might not natively support \UWMTemporaryAssumption{}/\DWMTemporaryAssumption{}, thus resulting in $\DuplicateAbbr{}$'s $A_1$ not looping on an output tuple $t_o$ on the basis of $t_o$ being a late arrival or forwarding late arrivals to $A_2$.
We thus introduce two algorithms to enforce \UWMTemporaryAssumption{}/\DWMTemporaryAssumption{}, defining four (non-concurrent) operations for SPEs to handle a tuple $t$ or a watermark $W$ of streams $S_E$ and $S_{A_2}$ (see \autoref{fig:loops}):
\begin{itemize}[leftmargin=*]
    \item $processT(t)$, run
    when $t$ is 
    fed to $S_E/S_{A_2}$
    \item $processW(W)$, run
    when $W$ is 
    fed to $S_E/S_{A_2}$,
    \item $forwardT(t)$, run
    to feed $t$ to the operators fed by $S_E/S_{A_2}$, and
    \item $forwardW(W)$, run
    to feed $W$ to the operators fed by $S_E/S_{A_2}$.
\end{itemize}
The algorithms make use of the following data structures:
\begin{itemize}[leftmargin=*]
    \item $Queue$ (FIFO), with associated methods $enq$ and $deq$ and use notation $Q[i]$ to refer to the element at $Q$'s index $i$, and 
    \item $TreeMap$, a map that allows for sorted traversal of its keys and associated values. We use the notation $m[k]$ to refer to the value of key $k$ in the $TreeMap$ $m$. Method $firstKey$ returns the value of the first key (without removing $k$ nor its value), while method $remove(k)$ removes key $k$ and the value associated to $k$. 
\end{itemize}

Listing~\ref{alg:uwm} presents the implementations of $processT$ and $processW$ to enforce \UWMTemporaryAssumption{} for $S_E$.
$S_E$ maintains four variables:
$B$ (\coderefoneline{alg:uwm}{alg:uwm:varB}), a bound on the highest watermark that can be forwarded by $S_E$,
$succ\Gamma$ (\coderefoneline{alg:uwm}{alg:uwm:varSuccGamma}), a $TreeMap$ that maintains, for the left boundary of each window instance $\window{}$ at $A_1$, the number of associated successor tuples,
$pendingW$ (\coderefoneline{alg:uwm}{alg:uwm:varPendingW}), a $Queue$ of watermarks that can be forwarded by $S_E$, and 
$L$ (\coderefoneline{alg:uwm}{alg:uwm:varL}), the Allowed Lateness of $A_1$.

Upon reception of a tuple $t$ (\coderefoneline{alg:uwm}{alg:uwm:processT}), 
$S_E$ forwards $t$ (\coderefoneline{alg:uwm}{alg:uwm:forward}) and 
proceeds 
updating $succ\Gamma$, increasing the entry for a window with left boundary $t.\tau$ (see Lemma~\ref{obs:tumbling}) with the number of successors carried by $t$ if $t$ comes from the $\EncapsulateAbbr$ operator preceding $A_1$ (i.e., if $t[2]=-1$, \coderefoneline{alg:uwm}{alg:uwm:tfromE}), or 
decreasing the entry at $t.\tau$ if $t$ is a successor of a previous tuple from $\EncapsulateAbbr$ (\coderefoneline{alg:uwm}{alg:uwm:decreaseSucc}).
If an entry in $succ\Gamma$ decreases to 0, such an entry is removed from $succ\Window$ (\codereftwolines{alg:uwm}{alg:uwm:removekeystart}{alg:uwm:removekeyend}).
$S_E$ proceeds then updating $B$, setting it to $\infty$ if all entries in $succ\Window$ have been cleared, or setting to the timestamp value of the left window boundary of the earliest successor yet to be received by $S_E$ (\codereftwolines{alg:uwm}{alg:uwm:updateBstart}{alg:uwm:updateBend}) plus $L$. 
Finally, $S_E$ tries to forward the latest watermark in $pendingW$ that is smaller than or equal to $B$, discarding any other earlier watermark (\codereftwolines{alg:uwm}{alg:uwm:forwardlateWstart}{alg:uwm:forwardlateWend}).

Upon reception of watermark $W$ (\coderefoneline{alg:uwm}{alg:uwm:processW}), $S_E$ immediately forwards $W$ if $W \leq B$ or stores it in $pendingW$, to later try to forward $W$ at a subsequent invocation of $processT$ (\codereftwolines{alg:uwm}{alg:uwm:checkWandB}{alg:uwm:storeW}).

Based on Listing~\ref{alg:uwm}, 
we note that.

\begin{lemma}
\label{thm:uwm}
If \watermarkDAssumption{} holds for $S_E$, then Listing~\ref{alg:uwm} enforces \UWMTemporaryAssumption{} for $A_1$.

\begin{algorithm}[h]
\footnotesize
\SetAlgoLined
\DontPrintSemicolon
\SetKwProg{Proc}{Function}{}{}
\BlankLine
$B=\infty$ \tcp{Bound on $W$ that $S_E$ can forward} \label{alg:uwm:varB}
$succ\Gamma{}$ \tcp{$TreeMap$ of $\window$'s right bound-\#successors} \label{alg:uwm:varSuccGamma}
$pendingW$ \tcp{$Queue$ of $W$} \label{alg:uwm:varPendingW}
$L$ \tcp{$A_1$'s Allowed Lateness} \label{alg:uwm:varL}
\BlankLine
\Proc{$processT(t)$}
{ \label{alg:uwm:processT}
    $forwardT(t)$ \tcp{Forward $t$} \label{alg:uwm:forward}
    \If{$t[2]=-1$}{\label{alg:uwm:tfromE}
        $succ\Gamma{}[t.\tau] \xleftarrow{} succ\Gamma{}[t.\tau] + |t[1]|$ \tcp{Keep track of succ. for $\window$ with left boundary $t.\tau$} \label{alg:uwm:updateSucc}
    } \Else {
        $succ\Gamma{}[t.\tau] \xleftarrow{} succ\Gamma{}[t.\tau] - 1$ \tcp{Decrease succ. for $\window$ with left boundary $t.\tau$} \label{alg:uwm:decreaseSucc}
        \If(\tcp*[h]{Got all succ. for $t.\tau$}){$succ\Window[t.\tau]=0$}{\label{alg:uwm:removekeystart}
            $succ\Window.remove(t.\tau)$\;\label{alg:uwm:removekeyend}
        }
    }
    \If(\tcp*[h]{Update $B$}){$|succ\Window|>0$}{\label{alg:uwm:updateBstart}
        $B \xleftarrow{} succ\Window.fistKey()+L$\;\label{alg:uwm:setBsuccessors}
    }\Else{
        $B \xleftarrow{} \infty$\;\label{alg:uwm:updateBend}
    }
    $nextW \xleftarrow{} -1$ \tcp{Forward latest $W|W\leq B$}\label{alg:uwm:forwardlateWstart}
    \While{$|pendingW|>0 \wedge pendingW[0]\leq B$}{
        $nextW \xleftarrow{} pendingW.deq()$\;    
    }
    \If{$nextW\neq-1$}{
        $forwardW(nextW)$\;\label{alg:uwm:forwardlateWend}
    }
}
\BlankLine
\Proc{$processW(W)$}
{ \label{alg:uwm:processW}
    \If(\tcp*[h]{If $W$ within $B$, send $W$}){$W \leq B$}{\label{alg:uwm:checkWandB}
        $forwardW(W)$\; \label{alg:uwm:forwardW1}
    } \Else(\tcp*[h]{Else, store $W$}) {
        $pendingW.enq(W)$\; \label{alg:uwm:storeW}
    }
}
\caption{Enforcing \UWMTemporaryAssumption{} for Stream $S_E$}
\label{alg:uwm}
\end{algorithm}
\end{lemma}

After discussing how \UWMTemporaryAssumption{} can be enforced for $S_E$, we now discuss how \DWMTemporaryAssumption{} can be enforced for $S_{A_2}$ based on Listing~\ref{alg:dwm}.

$S_{A_2}$ immediately forwards a tuple $t$ upon its reception and proceeds updating a variable $succ\Gamma$ (\codereftwolines{alg:dwm}{alg:dwm:forwardT}{alg:dwm:removekey}). Since, 
$S_{A_2}$ sees only the successors of a tuple $t'$ forwarded by $S_E$ to $A_1$, $S_{A_2}$ increases the corresponding entry for a tuple $t$ in  $succ\Gamma$ to $|t[1]|-1$ (since $t$ itself is one of the successors stemming from $t'[1]$).
Once $succ\Window$ is updated, $S_{A_2}$ forwards $t.\tau$ as a watermark if $succ\Window$ is empty, or the timestamp of the earliest left boundary of a window $- 1$ of a successor yet to be seen by $S_{A_2}$, if such a value is greater than that of the latest watermark forwarded by $S_{A_2}$ (\codereftwolines{alg:dwm}{alg:dwm:watermarkforwardstart}{alg:dwm:watermarkforwardend}).

Note that, being $T = succ(t)$, $T[0]$ is received by $S_{A_2}$ before a watermark $W|W\geq T[0].\tau$ (because $W$ is needed by $A_1$ to produce $T[0]$ and $W$ is fed to $S_{A_2}$ after $T[0]$, see \autoref{ssc:correctness}) while all other tuples in $T$ are received by $S_{A_2}$ after $W$. Hence, all tuples that result in increasing an entry of $succ\Window$ precede those decreasing an entry of $succ\Window$ and, if the earliest entry in $succ\Window$ decreases to $0$ upon the processing of $T[0]$, the processing of other tuples in $T$ can only decrease later entries in $succ\Window$.

Upon invocation of $processW$, $S_{A_2}$ immediately forwards $W$ if $succ\Window$ is empty, or the timestamp of the earliest left boundary of a window $- 1$ of a successor yet to be seen by $S_{A_2}$, if such a value is greater than that of the latest watermark forwarded by $S_{A_2}$ (\codereftwolines{alg:dwm}{alg:dwm:processW}{alg:dwm:end}).

Based on Listing~\ref{alg:dwm}, we note that:

\begin{lemma}
\label{thm:dwm}
If \watermarkDAssumption{} and \UWMTemporaryAssumption{} hold for $S_E$ and $A_1$, respectively, then Listing~\ref{alg:dwm} enforces \DWMTemporaryAssumption{} for $A_1$.

\begin{algorithm}[h]
\footnotesize
\SetAlgoLined
\DontPrintSemicolon
\SetKwProg{Proc}{Function}{}{}
\BlankLine
$succ\Gamma{}$ \tcp{$TreeMap$ of $\window$'s right bound-\#successors} \label{alg:dwm:varSuccGamma}
$lastW$ \tcp{Last $W$ forwarded by $S_{A_2}$}
\BlankLine
\Proc{$processT(t)$}
{ \label{alg:dwm:processT}
    $forwardT(t)$ \tcp{Forward $t$} \label{alg:dwm:forwardT}
    \If{$t[2]=-1$}{\label{alg:dwm:tfromE}
        $succ\Gamma{}[t.\tau] \xleftarrow{} succ\Gamma{}[t.\tau] + |t[1]|-1$ \tcp{Keep track of succ. for $\window$ with left boundary $t.\tau$, excluded $t$ itself.} \label{alg:dwm:updateSucc}
    } \Else {
        $succ\Gamma{}[t.\tau] \xleftarrow{} succ\Gamma{}[t.\tau] - 1$ \tcp{Decrease succ. for $\window$ with left boundary $t.\tau$} \label{alg:dwm:decreaseSucc}
        \If(\tcp*[h]{Got all succ. for $t.\tau$}){$succ\Window[t.\tau]=0$}{
            $succ\Window.remove(t.\tau)$\;\label{alg:dwm:removekey}
        }
    }
    \If(\tcp*[h]{Forward $t.\tau$ as watermark}){$|succ\Window|=0$}{\label{alg:dwm:watermarkforwardstart}
        $forwardW(t.\tau)$\;\label{alg:dwm:forwardW1}
        $lastW \xleftarrow{} t.\tau$\;
    }\ElseIf(\tcp*[h]{Forward watermark based on earliest pending succ.}){$succ\Window.firstKey()-1>lastW$}{
         $forwardW(succ\Window.firstKey()-1)$\;\label{alg:dwm:forwardW2}
        $lastW \xleftarrow{} succ\Window.firstKey()-1$\;\label{alg:dwm:watermarkforwardend}
    }
}
\BlankLine
\Proc{$processW(W)$}
{ \label{alg:dwm:processW}
    \If(\tcp*[h]{All pending succ. cleared}){$|succ\Window|=0$}{
        $forwardW(W)$\; \label{alg:dwm:forwardW3}
        $lastW \xleftarrow{} W$
    }\ElseIf(\tcp*[h]{Forward watermark based on earliest pending succ.}){$succ\Window.firstKey()-1>lastW$}{
        $forwardW(succ\Window.firstKey()-1)$\;\label{alg:dwm:forwardW4}
        $lastW \xleftarrow{} succ\Window.firstKey()-1$\;\label{alg:dwm:end}
    }
}
\caption{Enforcing \DWMTemporaryAssumption{} for Stream $S_{A_2}$}
\label{alg:dwm}
\end{algorithm}
\end{lemma}

\section{Relaxations and Extensions}
\label{sec:relax}

Before moving to our evaluation, we discuss in here how a relaxation on the requirement for $A$ to produce only zero or one tuples from a window instance $\window$ (see \autoref{sec:dsbasics}) can support the enforcing of $FM$ and $J$ semantics, and also show evidence about how $A$'s semantics can overlap, within and across SPEs and other frameworks, to other operators besides $F$, $M$, $FM$ and $J$ (main focus of this work).

\subsection{On the benefits of semantically richer $A$s}
\label{sec:aplus}

As discussed in \autoref{sec:native}, we introduced the \EncapsulateAbbr{} and \DuplicateAbbr{} operators to make up for $A$ constraint of producing only zero or one tuple from a window instance $\window$. Nonetheless, SPEs such as Flink allows users to produce an arbitrary number of output tuples from $\window$. By having proven $A$ suffices for enforcing $F$, $M$, $FM$, and $J$, we can conclude such a semantically richer $A$, which we refer to as $A^+$, can also enforce $F$, $M$, $FM$, and $J$ semantics.
Without a formal discussion for space reasons, it is sufficient to note that an $A^+$-based $\EncapsulateAbbr{}$ eliminates the need for $\DuplicateAbbr{}$, being able to immediately forward the tuples that would otherwise be embedded in $t_{\EncapsulateAbbr{}}[1]$ by \EncapsulateAbbr{} (see \autoref{sec:eandd}). Note that, if \DuplicateAbbr{} operators are not needed, \watermarkDAssumption{}, \cyclicGraphsAssumption{}, \UWMTemporaryAssumption{}, and \DWMTemporaryAssumption{}, can be lifted. 
We refer to \autoref{sec:evaluation} for an empirical comparison of $A$ and $A^+$ performance.

\subsection{Using $A$s to enforce custom semantics}
\label{sec:otherstate}

While common operators such as $F$, $M$, $FM$, and $J$ are our main focus, a question can be formulated on whether $A$s can also enforce the semantics of other operators, possibly user-defined/custom ones. While an efficient composition of $A$s might need to be discussed on a case-by-case basis, we believe such a question can be answered positively, since, as we show next, compositions of $A$s can be used to maintain states that go beyond those of time-based windows.

We define next a generic $O$ operator with an unbounded state (in terms of event time) so that each tuple $t$ fed to $O$ is processed exactly once and used to update a state based on all previously processed tuples. We note $O$'s state is unbounded in terms of the event time it spans but should be defined by the user not to have ever-growing space complexity.
As we show, $O$'s semantics can be enforced by relying on an $A$ with a sliding window and a loop to pour the state of a window instance $\window_l$ to that of $\window_{l+1}$. 
The user can then define $O$'s state in terms of a \textit{state tuple} that is carried across consecutive windows while accounting for $O$'s input tuples and $O$'s state tuples to possibly have different types. More concretely, we define $S_O = O(f_c,f_a,f_m,f_o,P,f_k,S_I)$ as a stateful operator so that:
\begin{description}
\item [Function $\bm{f_c}$] creates state tuple $t_s$ from input tuple $t_i$,
\item [Function $\bm{f_a}$] adds $t_i$ to an existing state tuple $t_s$, 
\item [Function $\bm{f_m}$] merges two state tuples $t_{s_1}$ and $t_{s_2}$,
\item [Function $\bm{f_o}$] produces an output tuple from $t_s$,
\item [Output period $\bm{P}$] is the period with which $f_o$ is invoked,
\item [Key-by function $\bm{f_k}$] is a key-by function (same as $A$), and 
\item [Input stream $\bm{S_I}$] is the input stream,
\end{description}

\noindent and state the following lemma:

\begin{lemma}
\label{thm:o}
The semantics of 
$S_O = O(f_c,f_a,f_m,f_o,P,f_k,S_I)$
can be enforced by composing $A$s as shown in Figure~\ref{fig:o} and Listing~\ref{alg:o} where \watermarkDAssumption{} holds for $S_{FM}$, \UWMTemporaryAssumption{},and \DWMTemporaryAssumption{}  hold for $A_1$, and $A_1$'s $L>D$.

\begin{figure}[h]
\includegraphics[width=\linewidth]{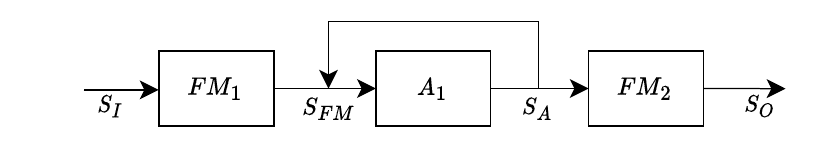}
\caption{Graph of $A$ operators implementing $O$'s semantics.}
\label{fig:o}
\end{figure}

\begin{algorithm}[h]
\footnotesize
\SetAlgoLined
\DontPrintSemicolon
\SetKwProg{Class}{Class}{}{}
\SetKwProg{Proc}{Function}{}{}
\BlankLine
\nonl $S_{FM} = FM(S_I,f_{FM})$, where: \tcp{\encircle{$FM_1$}}
\BlankLine
\Proc{$f_{FM}(t)$}
{
    \Return $\{t\} ^\frown \{\}$
}
\nonl\hrulefill\\
\nonl $S_A = A(\Window(P,P+\delta,S_{FM},f_k,L),f_o)$, where: \tcp{\encircle{$A_1$}}
\BlankLine
\Proc{$f_o(\window)$}
{
    $first\_t_s \xleftarrow{} True$\;
    \For(\tcp*[h]{For each $t \in \window$}){$t \in \window.\zeta$}{
        \If(\tcp*[h]{$t$ falls only in $\window$}){$t.\tau \neq \window.l+P-\delta$}
        {\label{alg:o:processonlyone}
            \If(\tcp*[h]{$t$ from $FM_1$}){$t[1] \neq \emptyset$}
            {
                \If(\tcp*[h]{state not created}){$first\_t_s$}
                {
                    $t_s \xleftarrow{} f_c(t[1])$\;\label{alg:o:createstate}
                    $first\_t_s \xleftarrow{} False$\;
                }
                \Else(\tcp*[h]{state created})
                {
                    $t_s \xleftarrow{} f_a(t_s,t[1])$
                }
            }
            \Else(\tcp*[h]{$t$ from $A_1$})
            {
                \If(\tcp*[h]{state not created}){$first\_t_s$}
                {
                    $t_s \xleftarrow{} t[2]$\;
                    $first\_t_s \xleftarrow{} False$\;
                }
                \Else(\tcp*[h]{state created})
                {
                    $t_s \xleftarrow{} f_m(t_s,t[2])$
                }
            }
        }
    }
    \Return $\{\} ^\frown \{t_s\}$
}
\nonl\hrulefill\\
\nonl $S_{O} = FM(S_A,f_{FM})$, where: \tcp{\encircle{$FM_2$}}
\BlankLine
\Proc{$f_{FM}(t)$}
{
    \Return $f_o(t[2])$
}
\caption{\small $A$ operators implementing $O$'s semantics.}
\label{alg:o}
\end{algorithm}
\end{lemma}

Note that, since $A_1$'s window instances are processed upon their expiration, the tuples within each instance can be sorted on their type to control the order in which $f_c$, $f_a$, and $f_m$ are invoked.

\section{Evaluation}
\label{sec:evaluation}

Our work demonstrates that common SPE operators can be expressed as a composition of $A$ operators. As a consequence, one may build an SPE that only adopts $A$ operators and use it as a reference to validate the semantics of \Dedicated{} operators, that is, operators that are offered by specific SPEs.  This section investigates the performance of such a reference SPE by comparing \AggBased{} and \Dedicated{} implementations of common operators.
The comparison does not aim to determine the best implementations, since \Dedicated{} implementations are expected to perform better.
Instead, we consider various operators, implementation characteristics, workloads, selectivity, and tuple processing costs, and we show that the performance of \AggBased{} operators is often comparable to \Dedicated{} ones. Thus, a reference SPE that relies on \AggBased{} operators can certainly be used for testing and validation purposes, and may even be adopted in practice in some scenarios.
To build such a reference SPE, we rely on Flink. We consider both Flink's $A$ and $A+$ (see \autoref{sec:aplus}) operators. We use the latter to shed light on how semantically-richer $A$ operators reduce the gap between \Dedicated{} and \AggBased{} implementations.
We also compare \Dedicated{} and \AggBased{} implementations choosing applications representatives of analysis deployed at opposite ends of the cloud-edge spectrum, thus accounting for the different computational power of the underlying hardware.

In the following, we present our setup in \autoref{sec:eval:setup} and we discuss results in \autoref{sec:eval:results}. We refer to \Dedicated{} implementations as $D$, and to \AggBased{} implementations as $A$ and $A+$.

\subsection{Setup}
\label{sec:eval:setup}

\noindent\textbf{Datasets.}
We use (1) a portion of the 43 million atomic Wikipedia edits from~\cite{WikiAtomicEdits} and (2) a dataset of 565 2D scans from a rangefinder sensor used in an industrial setup~\cite{mohamed2018detection,mohamed20192d}.
For (1), tuples' attributes include 
$\tau$ (\texttt{long}), set upon tuple forwarding, 
$orig$ (\texttt{String}), the original entry, 
$change$ (\texttt{String}), the text being added, and 
$updated$ (\texttt{String}), the modified entry.
For (2), tuples' attributes include 
$\tau$ (\texttt{long}), set upon tuple forwarding, 
$id$ (\texttt{int}), the tuple id, and 
$dist$ (array of \texttt{double}), the sensor distance readings.

\noindent\textbf{Experiments.} We compare $D$, $A$, $A+$ implementations of $FM$ and $J$ operators. Table~\ref{tab:eval-outline} lists the experiments we use to evaluate the performance of the various implementations.  For each experiment, we indicate the operator it involves, its selectivity, its computational cost, and the hardware device we use to execute it.

\noindent\textbf{Hardware.}
We use (1) an Intel Xeon E5-2637 v4 @ 3.50GHz (4 cores, 8 threads), and 64 GB RAM server to process Wikipedia edits, and (2) Odroid-XU4~\cite{OdroidXU42016a} (or simply \emph{Odroid}) edge devices~\cite{fuEdgeWiseBetterStream, palyvos-giannasHarenFrameworkAdHoc2019}, with Samsung Exynos5422 Cortex-A15 2Ghz and Cortex-A7 Octa core CPUs, and 2 GB RAM to process 2D scans.

\noindent\textbf{Software.}
We use Apache Flink 1.15.2~\cite{flink} and Ubuntu 18.04. Due to an open issue that possibly deadlocks loops in Flink~\cite{openissue}, we define our own loop-handling mechanism to meet \UWMTemporaryAssumption{} and \DWMTemporaryAssumption{}.

\noindent\textbf{Methodology.}
Performance is measured in terms of maximum sustainable throughput and latency.
$FM$'s throughput is measured in tuples/second (t/s) and $J$'s throughput is measured in comparisons/second (c/s). 
Latency, for both $FM$ and $J$, refers to the delay in the production of an output tuple $t_o$ after all the input tuples (one or more) that jointly result in $t_o$'s production have been processed by the operator producing $t_O$.
We consider as maximum sustainable throughput the one an SPE can sustain for 10~minutes without exceeding a latency of 15 seconds more than 3 times (we then say such an experiment is \textit{successful}).
We run each 10~minutes experiment for various injection rates. The first warm-up and the last cool-down minutes of each execution do not contribute to the results.
For throughput and injection rate, we report the average per-second value. For latency, we report the 99th percentile of the per-second value. Results are averaged over three repetitions.

We study varying selectivity and per-tuple processing costs. Selectivity for $FM$ is the average number of output tuples per input tuple. Selectivity for $J$ is the average number of tuples produced per comparison.
For $FM$, we consider a selectivity smaller than one for $FM$ to resemble an $F$ and a selectivity equal to one for $FM$ to resemble an $M$.
Since $J$'s comparisons are quadratic in the input rate~\cite{gulisano2017performance}, we consider selectivity values smaller than $1$ as in related work~\cite{teubner2011soccer,gulisano2017performance}.
\autoref{tab:eval-outline} overviews all the experiments discussed next.

\begin{table*}[ht!]
\scriptsize
\centering
\begin{threeparttable}
\caption{List of experiments.}
\label{tab:eval-outline}
\begin{tabularx}{\textwidth}{p{0.015\linewidth}p{0.1\linewidth}p{0.018\linewidth}p{0.02\linewidth}p{0.232\linewidth}
p{0.015\linewidth}p{0.1\linewidth}p{0.018\linewidth}p{0.02\linewidth}p{0.232\linewidth}}
\toprule
\multicolumn{5}{c}{High-end server (upper-case ID)} & \multicolumn{5}{c}{Odroid device (lower-case ID)} \\
\cmidrule(lr){1-5} \cmidrule(lr){6-10}
\textbf{ID} & \textbf{Selectivity} & \textbf{Cost} & \textbf{Op.} & \textbf{Notes} & 
\textbf{ID} & \textbf{Selectivity} & \textbf{Cost} & \textbf{Op.} & \textbf{Notes} \\ 
\midrule
\texttt{LLF} & Low ($\sim 5e^{-3}$) & Low & $FM$ & Find most frequent word in $orig$. Forward if length $>$ 10 chars & 
\texttt{llf} & Low ($0.2$) & Low & $FM$ & Convert coordinates from polar to Cartesian. Forward if avg dist. $>$ 3m \\
\texttt{ALF} & Avg ($1$) & Low & $FM$ & Find the most frequent word in $orig$ and forward it & 
\texttt{alf} & Avg ($1$) & Low & $FM$ & Convert coordinates from polar to Cartesian\\
\texttt{HLF} & High ($3$) & Low & $FM$ & Find top-3 frequent words in $orig$. Forward as separate tuples &
\texttt{hlf} & High ($3$) & Low & $FM$ & Convert coordinates from polar to Cartesian, and split/forward in $3$ parts \\
\texttt{LHF} & Low ($\sim 3e^{-4}$) & High & $FM$ & Find most frequent word in $orig$, $change$, and $update$. Forward them in a single tuple if all their lengths are $>$ 10 chars & 
\texttt{lhf} & Low ($\sim 0.7$) & High & $FM$ & Convert coordinates from polar to Cartesian from reference point. Forward if avg dist. $>$ 3m\\
\texttt{AHF} & Avg ($1$) & High & $FM$ & Find most frequent word in $orig$, $change$, and $update$. Forward them in a single tuple &
\texttt{ahf} & Avg ($1$) & High & $FM$ & Convert coordinates from polar to Cartesian from reference point\\
\texttt{HHF} & High ($\sim 2.3$) & High & $FM$ & Find top-3 frequent words in $orig$, $change$, and $update$. Forward as separate triplets &
\texttt{hhf} & High ($3$) & High & $FM$ & Convert coordinates from polar to Cartesian from reference point, and split/forward in 3 parts\\
\texttt{LLJ} & Low ($\sim 1e^{-4}$) & Low & $J$ & Match two distinct (case insens.) $orig$ if they have the same length and if $|orig|>$ 210 chars. Key-by number of words in $change$, \WA=1s and \WS=3s &
\texttt{llj} & Low ($\sim 8e^{-5}$) & Low & $J$ & Match two distinct scans if the sum diffs in $dist$ is $<$ 0.5m. \WA=0.5s and \WS=1s\\
\texttt{ALJ} & Avg ($\sim 1e^{-3}$) & Low & $J$ & As \texttt{LLJ}, but $|orig|>150$ &
\texttt{alj} & Avg ($\sim 8e^{-4}$) & Low & $J$ & As \texttt{llj} but sum diffs $<$ 0.6m\\
\texttt{HLJ} & High ($\sim 3e^{-3}$) & Low & $J$ & As \texttt{LLJ}, but $|orig|>100$ &
\texttt{hlj} & High ($\sim 5e^{-3}$) & Low & $J$ & As \texttt{llj}, but sum diffs $<$ 0.7m\\
\texttt{LHJ} & Low ($\sim 1e^{-4}$) & High & $J$ & As \texttt{LLJ}, but \WS=10s &
\texttt{lhj} & Low ($\sim 6e^{-5}$) & High & $J$ & As \texttt{llj}, but \WS=2s\\
\texttt{AHJ} & Avg ($\sim 1e^{-3}$) & High & $J$ & As \texttt{LLJ}, but $|orig|>150$ and \WS=10s &
\texttt{ahj} & Avg ($\sim 7e^{-4}$) & High & $J$ & As \texttt{llj}, but sum diffs $<$ 0.6m and \WS=2s\\
\texttt{HHJ} & High ($\sim3e^{-3}$) & High & $J$ & As \texttt{LLJ}, but $|orig|>100$ and \WS=10s &
\texttt{hhj} & High ($\sim3e^{-3}$) & High & $J$ & As \texttt{llj}, but sum diffs $<$ 0.7m and \WS=2s\\
\bottomrule
\end{tabularx}
\end{threeparttable}
\end{table*}

\subsection{Results}
\label{sec:eval:results}

We now present the results of our evaluation, first considering the $FM$ operator and then the $J$ operator.

\subsubsection*{$FM$ operator}

\begin{figure}[h]
\centering
\includegraphics[width=0.8\linewidth]{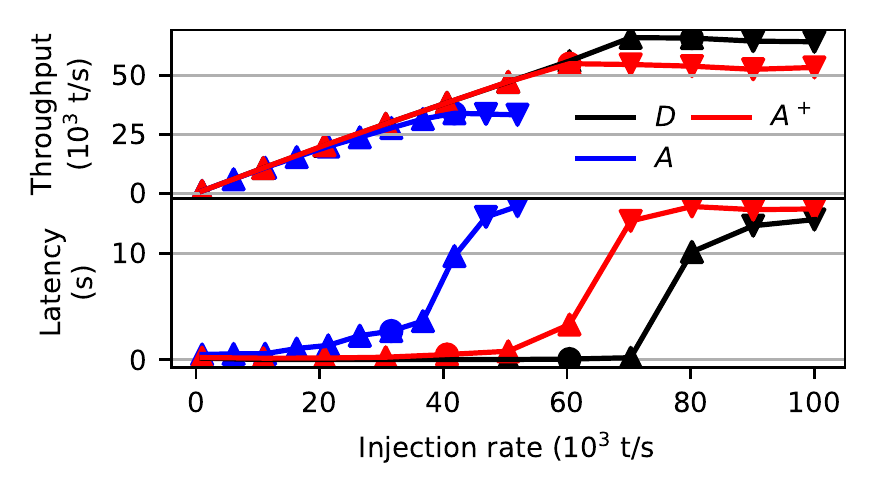}
\caption{\texttt{AHF} - Throughput/latency vs. injection rate}
\label{fig:FM:scalability}
\end{figure}

We begin by showing the results for one specific experiment and repetition, noting similar behavior is observed for other experiments. 
Figure~\ref{fig:FM:scalability} shows how throughput (top) and latency (bottom) change for an increasing injection rate (t/s) for the \texttt{AHF} experiment (see \autoref{tab:eval-outline}). 
Up-pointing markers indicate successful experiments, down-pointing markers unsuccessful ones.

The throughput initially increases linearly with the injection rate in all implementations but plateaus once the maximum sustainable throughput is reached, with the corresponding increase in latency, which steepens as the operator's maximum sustainable throughput is approached.
$D$'s maximum throughput is higher than that of $A/A^+$, although both remain in the same order of magnitude, with a degradation of approximately 50\% for $A$ and 16\% for $A^+$. 
For sustainable injection rates, we observe $A$'s latency grows faster than $A^+$'s, which in turn grows faster than $D$'s. This is because $D$ does not require watermarks to trigger the production of results (since $FM$ is a stateless operator) while $A$ and $A^+$ do, thus implying their latency is a function of the periodicity with which watermarks are forwarded by ingresses. 
Furthermore, for $A$, the latency increase is also due to the delay in the forwarding of watermarks paid to enforce \UWMTemporaryAssumption{} and \DWMTemporaryAssumption{} (see \autoref{sec:loops}). 
$A^+$'s higher throughput and lower latency (compared to $A$) is also due to $A^+$ not relying on the \DuplicateAbbr{} operator (see~\autoref{sec:aplus}) and thus not requiring loops.

\begin{figure}[h]
\centering
\includegraphics[width=0.48\linewidth]{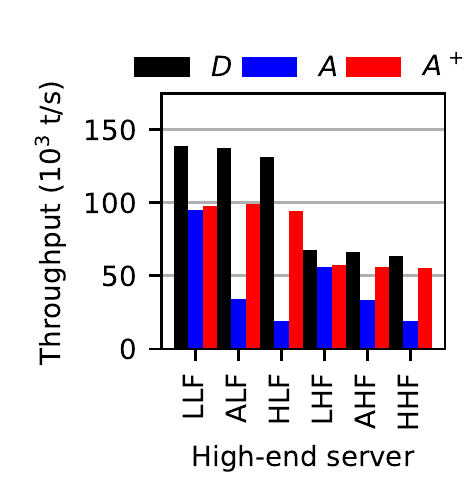}
\includegraphics[width=0.48\linewidth]{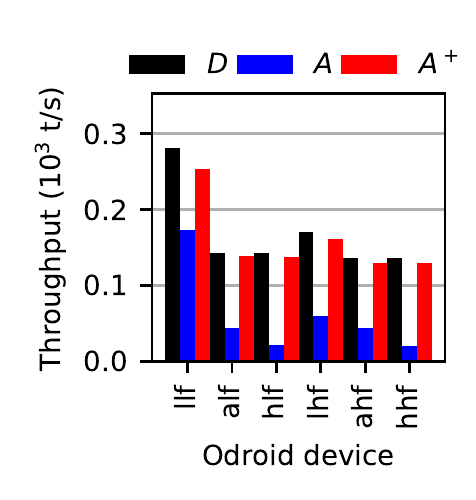}
\caption{$FM$ - Average throughput for all experiments.}
\label{fig:FM:bar:throughput}
\end{figure}

\begin{figure}[h]
\centering
\includegraphics[width=0.48\linewidth]{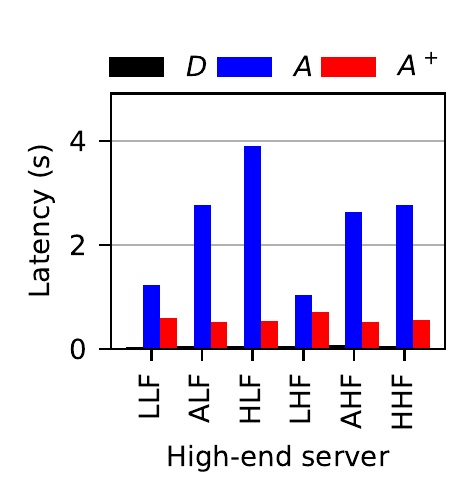}
\includegraphics[width=0.48\linewidth]{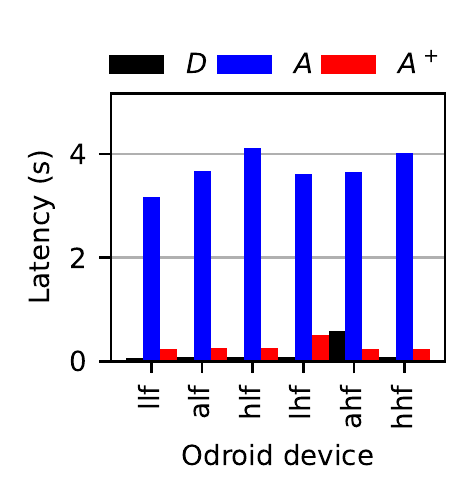}
\caption{$FM$ - Average latency for all experiments.}
\label{fig:FM:bar:latency}
\end{figure}

Figures~\ref{fig:FM:bar:throughput} and \ref{fig:FM:bar:latency} contain a comparison of throughput and latency across all experiments.
The value of throughput (Figure~\ref{fig:FM:bar:throughput}) for $D$ is not significantly affected by selectivity, but it is affected by per-tuple computational cost.
This does not hold for $A$, though: when $FM$'s selectivity is close to $0$, the two $A$ operators of \DuplicateAbbr{} (see~\autoref{sec:loops}) are only processing a few tuples. 
Hence, while the $A$ defining $\EncapsulateAbbr_{FM}$ is not as efficient as $D$, $\EncapsulateAbbr_{FM}$ and $D$'s performance are in the same order of magnitude (e.g., 34\% less throughput for $A$ in \texttt{LLF}).
When the rate fed to \DuplicateAbbr{} grows, the throughput drops, down to e.g., a 90\% for \texttt{HLF}.
Further evidence of \DuplicateAbbr{} impact on $A$'s performance is given by $A^+$'s performance, for which selectivity does not affect throughput (as for $D$) and thus results in a higher maximum sustainable throughput, with negligible differences from $D$ in the Odroid-based experiments.

For the latency metric, we can note that both $A$ and $A^+$ consistently exhibit higher latency than $D$, as expected. However, we note that the difference in latency between $A$ and $D$ becomes smaller when the analysis is run on an edge device such as an Odroid. 
We also note that $A$ exhibits a steady increase in latency as selectivity increases. This can be attributed to the looping of tuples within the \DuplicateAbbr{} operator.
The growing trend in latency is also present for $D$ and $A^+$ when executed on an Odroid. However, this trend is more pronounced on the high-end server. On the latter, $D$'s latency is orders of magnitude lower than that of $A$ and $A^+$. This is consistent with our expectation that $D$'s evaluation is based on a single stateless operator.
Finally, note that $A^+$'s latency remains in the sub-second range even on high-end servers. This latency is acceptable for many real-world streaming applications.

\subsubsection*{$J$ operator}

\begin{figure}[h]
\centering
\includegraphics[width=0.8\linewidth]{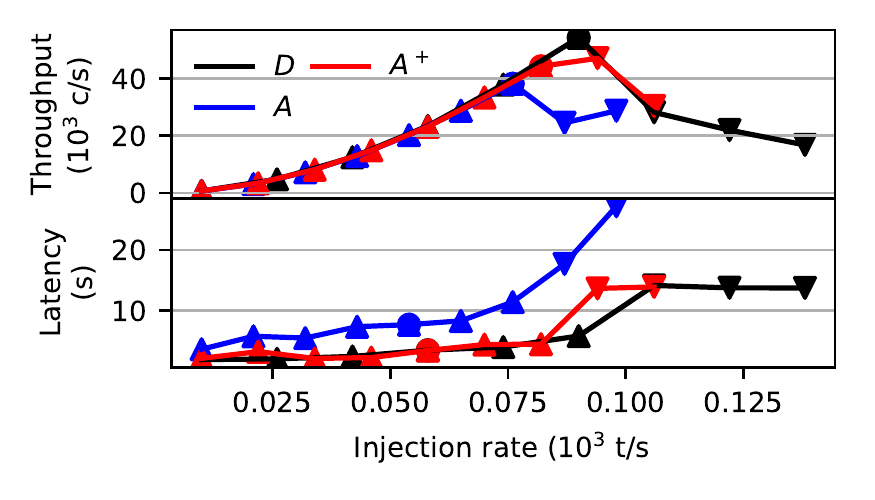}
\caption{\texttt{ahj} - Throughput/latency vs. injection rate.}
\label{fig:J:scalability}
\end{figure}

\autoref{fig:J:scalability} shows the throughput and latency for the $J$ operator in experiment \texttt{ahj}.
In this case, the behaviors of $D$ and $A^+$ implementations are close, while the latency of $A$ grows faster as the rate increases. For $A$ and $A^+$, all the comparisons for a given window $\window$ are done at once when $\window$'s right boundary falls before $A$'s watermark, while for the $D$ they are done as tuples are coming in. Since $A$ also requires the subsequent unfolding from \DuplicateAbbr{}, it results in higher amounts of work to be carried out on window expiration and thus lower throughput/higher latency.

\begin{figure}[h]
\centering
\includegraphics[width=0.48\linewidth]{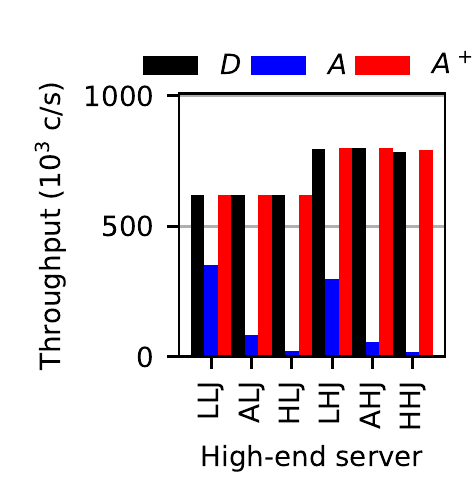}
\includegraphics[width=0.48\linewidth]{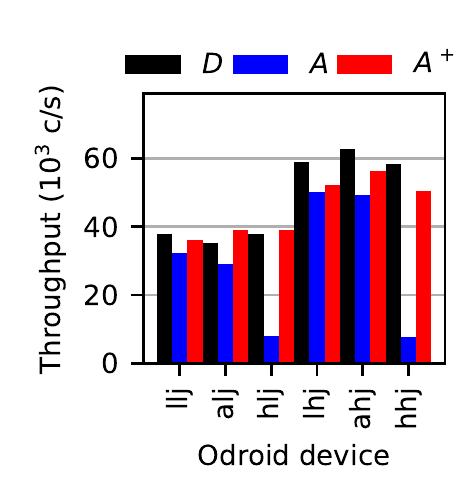}
\caption{$J$ - Average throughput for all experiments.}
\label{fig:J:bar:throughput}
\end{figure}

\begin{figure}[h]
\centering
\includegraphics[width=0.48\linewidth]{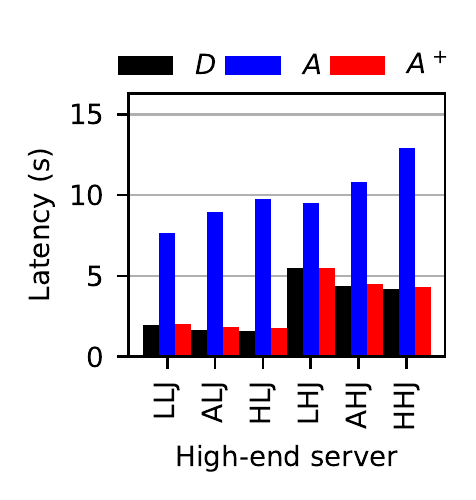}
\includegraphics[width=0.48\linewidth]{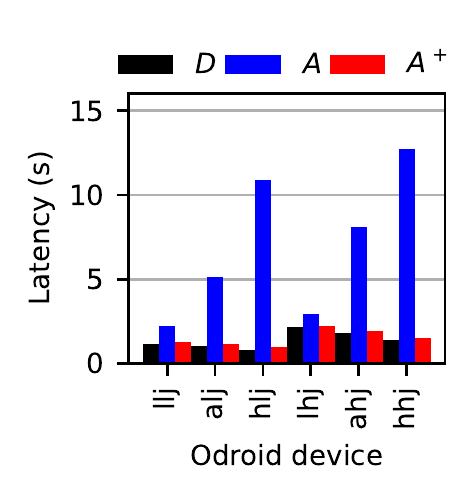}
\caption{$J$ - Average latency for all experiments.}
\label{fig:J:bar:latency}
\end{figure}

Figures~\ref{fig:J:bar:throughput} and \ref{fig:J:bar:latency} contain a comparison of throughput and latency for all experiments. The trends are similar to those of $FM$, with minor differences in terms of absolute values. In particular, both $A^+$ and $J$ operators show negligible differences. In some cases, we also note minor differences between $A$ and $D$, especially when $A$ and $D$ run on the Odroid device, such as in the case of the \texttt{llj} experiment.
We also observe that the growing trend of latency is mainly observed for the $A$ operator, while in this case that $A^+$ and $D$ exhibit comparable latency across all experiments.

\subsubsection*{Summary}
We compared throughput and latency performance for various experiments from two real-world applications, using also hardware representative of opposite ends of the computing spectrum power found in modern cyber-physical systems.

As expected, \Dedicated{} operators perform better. However, \AggBased{} implementations based on minimal $A$ operators or semantically-richer $A^+$ operators can often lead to comparable results, especially for stateful analysis such as that of $J$, in which both $D$ and $A$/$A^+$ need to rely on watermarks for the progression of their analysis and production of results.
We believe that, based on how much the performance gap narrows by relying on $A^+$ instead of $A$, an even semantically richer $A$ that could e.g., also produce intermediate results rather than only results computed on the expiration of a window instance, could further narrow such gap, something we plan to explore in future work.

\section{Related Work}
\label{sec:related}

The Dataflow model~\cite{akidau2015dataflow} considered here is the de-facto standard in modern distributed stream processing engines (SPEs).

The core ideas of the Dataflow model may already be found in the seminal MapReduce~\cite{dean:CACM:2008:mapreduce} data processing system.  In the early 2000s, MapReduce has shown the benefits of a programming abstraction where the developers only need to define the computation carried out by individual operators, with the system runtime handling all the complexity related to concurrent and distributed execution, such as deployment, scheduling, communication, concurrency control, and fault tolerance.
Around the same years, early SPEs like Aurora~\cite{abadi2003aurora} already built on a Dataflow programming and execution model.

Today, Dataflow is widely adopted in distributed data processing systems~\cite{margara2022dataintensive}, e.g., by systems such as Apache
Spark~\cite{zaharia:CACM:2016:spark}, Apache
Flink~\cite{carbone:IEEEB:2015:flink}, Apache
Storm~\cite{toshniwal:SIGMOD:2014:Storm} and
Heron~\cite{kulkarni:SIGMOD:2015:Twitter_Heron}, Kafka
Stream~\cite{bejeck:2018:KafkaStreams}, and
Samza~\cite{noghabi:VLDB:2017:Samza}, just to name a few.
It also powers the execution of analytical queries within some distributed database systems, such as F1~\cite{shute:VLDB:2013:F1}.
Core concepts such as time-based windows, needed for our minimalistic Aggregate operator (see \autoref{sec:dsbasics}) are found in relational/SQL streaming~\cite{begoli2019one}, RDF streaming, stream reasoning platforms~\cite{margara:JWS:2014:streaming}, and even spreadsheet frameworks such as ActiveSheets~\cite{hirzel2018stream}.

We are not aware of previous work studying the relationship between operators in Dataflow-based SPEs and seeking a core set of operators to express queries, as we do here. However, there are related areas of investigation that may inspire future research.

Work such as~\cite{soule2010universal} defines a calculus for streaming applications and shows how higher-level languages like CQL~\cite{arasu2006cql} can be ported to it. Note, though, \cite{soule2010universal} encapsulates semantics in orthogonal opaque functions, while we explicitly rely on basic Dataflow constructs to show semantic overlap within and across SPEs.

Some works focus on the semantics of individual aspects of SPEs.  For instance, the SECRET model provides a minimum set of parameters to precisely define windows~\cite{botan:VLDB:2010:SECRET}, which can be applied to identify possible differences across systems~\cite{affetti2017defining}.  Akidau et al.~\cite{akidauwatermarks} study the semantics of watermarks and the differences between two system implementations.  G\'{e}vay et al.~\cite{gevay:CSur:2022:iterations} survey and analyze different approaches to handle iterations in Dataflow systems.
Other works focus on higher-level abstractions on top of Dataflow-based SPEs e.g., through abstractions for relational data processing, exploiting the duality of data streams and relational tables~\cite{sax:BIRTE:2018:StreamsTables}.  Fernandez et al.~\cite{fernandez:ATC:2014:MakingStateExplicit} propose an object-oriented programming model that can be automatically translated to a Dataflow graph.
Yet other works study how to extend the expressivity of Dataflow-based SPEs. Naiad~\cite{murray:SOSP:2013:Naiad} offers an extended programming model to control how stream elements traverse nested loops through explicit vector timestamps.  CIEL~\cite{murray:NSDI:2011:CIEL} provide primitives to dynamically instantiate new operators, thus allowing for runtime definition of the graph of computation, for instance, based on the value of data.  
There are also various proposals to enrich the Dataflow model with some form of mutable state.  
TSpoon~\cite{affetti:JPDC:2020:tspoon} provides primitives to query and access the state of operators, and it lets developers specify portions of the graph of computation that need to be accessed and updated with transactional semantics.  
Symmetrically, S-Store~\cite{cetintemel:VLDB:2014:sstore} implements an SPE within a relational database core, thus offering transactional semantics for the internal state of stateful operators.

Alternative programming models for stream processing exist~\cite{cugola:CSur:2012:survey}.  In particular, Complex Event Recognition (CER)~\cite{giatrakos:VLDBJ:2020:CER} considers stream elements as occurrences of events and aims to recognize \emph{patterns} of such events.  Interestingly, the semantics of CER operators have been studied in detail~\cite{grez:TODS:2021:Formal}: our paper aims to provide a similar contribution to the Dataflow model.
In the context of CER, the equivalence between operators has been exploited to rewrite queries to an equivalent form that enables optimized execution~\cite{shchultz:DEBS:2009:rewriting}.  We believe that a deeper understanding of the semantic relations between operators may enable similar strategies for the Dataflow model, complementing existing optimization strategies~\cite{Hirzel2014CSP}, and potentially exploiting higher-level definitions of queries~\cite{guerriero:TOSEM:2021:StreamGen}.

\section{Conclusions}
\label{sec:conclusions}

Our work aimed to identify a core set of operators, common to all SPEs based on the DataFlow model, which suffices to enforce the semantics of common stream processing pipelines. Our formal analysis revealed that a single minimalistic Aggregate operator is sufficient to enforce the semantics of all common stateless and stateful operators defined in the DataFlow model. This finding implies that streaming applications within one SPE can be easily ported to other SPEs or Big Data frameworks able to support such a minimal Aggregate operator. Additionally, we found that the performance difference between the Aggregate operator and other streaming operators can be minimal in various real-world scenarios.

We also revealed that an Aggregate operator can be used to define stateful operators beyond common window-based ones. We plan to expand on this observation in future work, also exploring how the performance of streaming applications based on compositions of Aggregate operators evolve in distributed/parallel deployments.

\appendix
\section{Watermarks for an $A$ with a Loop}
\label{app:A}

We here elaborate on assumptions \UWMTemporaryAssumption{} and \DWMTemporaryAssumption{} (see \autoref{sec:smps}) about the handling of watermarks and looping tuples, discussing what could happen if an SPE maintains an $A$'s watermark based on the DataFlow model but without extra actions based on $A$'s being fed by a loop.

Let us assume 
$A$ maintains, for key $k$, a window instance $\window$ covering $[\text{10:00}-\text{11:00})$.
For $\window$ to result in an output tuple $t_o$, $\watermarkof{A}$ must grow to $\text{11:00}$ or more. If $W^j$ carrying time $\text{12:00}$ is the watermark triggering the production of $t_o$, and $t_o$ is fed again to $A$, $t_o$ will constitute a late arrival based on $t_o.\tau$, i.e., 11:00-$\delta$ and $\watermarkof{A}$, i.e., 12:00 (see \autoref{sec:dsbasics}). 
In general, observing that, because of \watermarkDAssumption{}, $t_o \in trig(W^i)$ can carry timestamps in 
$[W^j-D,W^j)$,
setting $A$'s allowed lateness to $L\geq D$, could prevent $t_o$ from not being processed by $A$.

Such a precaution is not sufficient, though. Without additional assumptions on the order with which $A$ interleaves its input tuples/watermarks (including the tuples from a loop), one cannot exclude that, upon reception of $t_o$, $A$'s processing of watermarks from its upstream peers resulted in $\watermarkof{A}$ to grow to an arbitrary value. If $L=D$ and $D=1$ hour and if, upon the reception of $t_o$, $\watermarkof{A}$ grew to e.g., $\text{12:05}$, then $t_o$ would still not be processed by $A$.

Finally, it also note that, even if the arbitrary interleaving of input tuples/watermarks at $A$ does not result in a $t_o$ fed through $A$'s loop not being processed, the forwarding of $W^i$ from $A$ to its stateful downstream peers (see a\autoref{ssc:correctness}) would require the latter to handle late arrivals through their Allowed Lateness (see \autoref{sec:latearrivals}).

\section{Proofs of Lemmas}
\label{app:proofs}


\begin{proof}
(Lemma~\ref{obs:tumbling})
The left boundary $\window.l$ of the window instance to which $t$ contributes to can be computed as $\floor{\nicefrac{t.\tau}{\WA}}\WA$ for a tumbling window. If $\WA=\delta$, then $\floor{\nicefrac{t.\tau}{\WA}}=\nicefrac{t.\tau}{\WA}$ and thus $\window.l=t.\tau$. Moreover, $t_o.\tau$ is set to $\window.l+\WS-\delta=t_i.\tau$.
\end{proof}

\begin{proof}
(Lemma~\ref{thm:noduplicates})
By contradiction; let us assume 
$t^l_o$ and $t^m_o$ are produced by $A_1$ and are so that $t^l_o=t^m_o$.
Noting that any tuple produced by $A_1$ for a window instance $\window$ shares the same timestamp of the tuples that fall in $\window$ (see Lemma~\ref{obs:tumbling}), and noting $A_1$ does not alter $t_o[1]$ if $t_o[2]\neq-1$ (\codereftwolines{alg:duplicate}{alg:duplicate:m2:checkindex}{alg:duplicate:m2:increase}),
the existence of $t^l_o$ and $t^m_o$ implies the existence of $t^{l'}_o$ and $t^{m'}_o$ produced by $A_1$ so that $t^{l'}_o[0:1]=t^l_o[0:1]$, $t^{l'}_o[2]=0$, $t^{m'}_o[0:1]=t^m_o[0:1]$, and $t^{m'}_o[2]=0$ (\codereftwolines{alg:duplicate}{alg:duplicate:m2:iffromE}{alg:duplicate:m2:iffromEend}).
Assuming $t^{l'}_o$ and $t^{m'}_o$ are produced by $A_1$ processing two identical tuples $t^{l}_i$ and $t^{m}_i$ from $S_E$ leads to a contradiction, because if $t^{l}_i=t^{m}_i$ then both fall in the same window instance $\window$ and $\window$ results only in one output tuple.
Assuming $t^{l'}_o$ and $t^{m'}_o$ are produced by $A_1$ processing two or more input tuples $t^{l_1}_i,t^{l_2}_i,\ldots$ and $t^{m_1}_i,t^{m_2}_i,\ldots$ so that $t^{l_j}_i\neq t^{m_j}_i, \forall l_j,m_j$ leads also to a contradiction because $t^{l_j}_i\neq t^{m_j}_i$ implies that the concatenations $t^{l_1}_i[1]^\frown t^{l_2}_i[1]^\frown \ldots$ and $t^{m_1}_i[1]^\frown t^{m_2}_i[1]^\frown \ldots$ differ too, and thus that $t^{l'}_o \neq t^{m'}_o$.
\end{proof}

\begin{proof}
\textbf{(Lemma~\ref{thm:uwm})}
A watermark $W$ is always forwarded under the condition that $W \leq B$ (\codereftwolinesDouble{alg:uwm}{alg:uwm:forwardlateWstart}{alg:uwm:forwardlateWend}{alg:uwm:checkWandB}{alg:uwm:forwardW1}).

If $B=\infty$, then $W$ is a watermark received before any tuple $t$ since $B$ is initialized at $\infty$, or $succ\Window$ is empty. In the first case, there is no pending tuple from $S_{A_2}$ yet to be processed. In the latter, any tuple $t$ from $E$ that increased $succ\Gamma[t.\tau]$ upon invocation of $processT(t)$ has been followed by all $t' \in succ(t)$ that, upon their related invocation of $processT(t')$, decreased $succ\Gamma[t'.\tau]$ to $0$. Hence, there is no tuple yet to be processed by $A_1$.

If $B\neq \infty$, then the distance between $W$ and the earliest tuple $t'$ from $A_1$ yet to be processed by $A_1$ is smaller than or equal to $L$. When $t'$ is received at $A_1$ after $W$, the condition $t'.\tau + \delta > \watermarkof{A_1}+L$ will be met (note that based on Lemma~\ref{obs:tumbling} $t'$ falls in a window instance whose left boundary is equal to $t'.\tau$), since it implies $\watermarkof{A_1}<t'.\tau+\delta+L$, and the latest watermark $W$ fed to $A_1$ by $S_{E}$ is so that $W \leq t'.\tau+L$.
\end{proof}

\begin{proof}
\textbf{(Lemma~\ref{thm:dwm})} 
If $W$ is forwarded upon its reception (\coderefoneline{alg:dwm}{alg:dwm:forwardW3}), then $|succ\Window|=0$. Hence, $\forall t \in succ(trig(W'))$ so that $W'<W$, $t$ was fed to $A_2$ before $W$.
All other invocations of $forwardW$ feed $A_2$ a watermark equal to 
(\coderefoneline{alg:dwm}{alg:dwm:forwardW1}) 
or smaller than 
(\coderefonelineDouble{alg:dwm}{alg:dwm:forwardW2}{alg:dwm:forwardW4}) a timestamp $\tau$ of an entry in $succ\Window$ that is not preceded by other entries with a count greater than 0. Hence, $\tau$ is fed as watermark to $A_2$ only after any $t \in succ(trig(W'))$ so that $W'\leq \tau$.
\end{proof}

\begin{proof}
\textbf{(Lemma~\ref{thm:o})} 
In \autoref{fig:o}, the stateless $FM_1$ defines a common set of attributes for the tuples fed to $A_1$, be they tuples from $S_I$ or $S_A$.
The core functionality is then run by $A_1$.
Based on the $\Window$ defined by $A_1$ only two consecutive window instances $\window_l = [lP,lP+P+\delta)$ and $\window_{l+1} = [(l+1)P,(l+1)P+P+\delta)$ overlap on $[(l+1)P,(l+1)P+\delta)$.
Nonetheless, tuples falling in the latter interval are only processed when in $\window_{l+1}$ (\coderefoneline{alg:o}{alg:o:processonlyone}).
Hence, $A_1$ processes every tuple exactly once.
We also note that, as soon as the very first tuple $t$ fed to $A_1$ is processed, a state tuple $t_s$ is created invoking $f_c$ (\coderefoneline{alg:o}{alg:o:createstate}).
Let $\window_l$ be the first window in which $t$ is processed. The resulting $t_s$ will have $t_s.\tau=(l+1)P$ and will be then processed within the window instance $\window_{l+1}$, since \watermarkDAssumption{}, \UWMTemporaryAssumption{}, and \DWMTemporaryAssumption{} prevent $t_s$ from not being processed on the basis of being a late arrival.
Iteratively, $t_s$ carries a state based on all processed tuples (not only within an individual window instance) from $\window_l$ to all the subsequent windows.
Finally, $FM_2$ runs $f_o$ on a state tuple with periodicity $P$, since $P$ is $A_1$'s $\WA$, thus enforcing $O$'s semantics.
\end{proof}

\bibliographystyle{IEEEtran}
\bibliography{refs}

\begin{thebibliography}{10}
\providecommand{\url}[1]{#1}
\csname url@samestyle\endcsname
\providecommand{\newblock}{\relax}
\providecommand{\bibinfo}[2]{#2}
\providecommand{\BIBentrySTDinterwordspacing}{\spaceskip=0pt\relax}
\providecommand{\BIBentryALTinterwordstretchfactor}{4}
\providecommand{\BIBentryALTinterwordspacing}{\spaceskip=\fontdimen2\font plus
\BIBentryALTinterwordstretchfactor\fontdimen3\font minus
  \fontdimen4\font\relax}
\providecommand{\BIBforeignlanguage}[2]{{%
\expandafter\ifx\csname l@#1\endcsname\relax
\typeout{** WARNING: IEEEtran.bst: No hyphenation pattern has been}%
\typeout{** loaded for the language `#1'. Using the pattern for}%
\typeout{** the default language instead.}%
\else
\language=\csname l@#1\endcsname
\fi
#2}}
\providecommand{\BIBdecl}{\relax}
\BIBdecl

\bibitem{akidau2015dataflow}
T.~Akidau, R.~Bradshaw, C.~Chambers, S.~Chernyak, R.~J.
  Fern{\'a}ndez-Moctezuma, R.~Lax, S.~McVeety, D.~Mills, F.~Perry, E.~Schmidt,
  and S.~Wittle, ``The dataflow model: a practical approach to balancing
  correctness, latency, and cost in massive-scale, unbounded, out-of-order data
  processing,'' \emph{Proc. Endowment}, vol.~8, no.~12, pp. 1792--1803, 2015.

\bibitem{flink}
``{Apache Flink},'' \url{https://flink.apache.org}, accessed:2023-1-27.

\bibitem{storm}
``{Apache Storm},'' \url{http://storm.apache.org}, accessed:2019-3-1.

\bibitem{liebre}
{Liebre SPE}, ``https://github.com/vincenzo-gulisano/liebre,'' 2022,
  accessed:2022-6-27.

\bibitem{beam}
``{Apache Beam},'' \url{https://beam.apache.org/}, accessed:2020-11-12.

\bibitem{gulisano2020role}
V.~Gulisano, D.~Palyvos-Giannas, B.~Havers, and M.~Papatriantafilou, ``The role
  of event-time order in data streaming analysis,'' in \emph{Proceedings of the
  14th ACM International Conference on Distributed and Event-based Systems},
  2020, pp. 214--217.

\bibitem{cederman2013concurrent}
V.~Gulisano, Y.~Nikolakopoulos, D.~Cederman, M.~Papatriantafilou, and
  P.~Tsigas, ``Efficient data streaming multiway aggregation through concurrent
  algorithmic designs and new abstract data types,'' \emph{ACM Trans. Parallel
  Comput.}, vol.~4, no.~2, pp. 11:1--11:28, Oct. 2017.

\bibitem{scalejoin}
V.~Gulisano, Y.~Nikolakopoulos, M.~Papatriantafilou, and P.~Tsigas,
  ``{ScaleJoin}: a deterministic, disjoint-parallel and skew-resilient stream
  join,'' \emph{IEEE Trans. Big Data}, vol.~7, no.~2, pp. 299--312, 2021.

\bibitem{teubner2011soccer}
J.~Teubner and R.~Mueller, ``{How soccer players would do stream joins},'' in
  \emph{Proc. of the 2011 ACM SIGMOD Int'l Conf. on Management of data}, 2011.

\bibitem{gulisano2022stretch}
V.~Gulisano, H.~Najdataei, Y.~Nikolakopoulos, A.~V. Papadopoulos,
  M.~Papatriantafilou, and P.~Tsigas, ``Stretch: Virtual shared-nothing
  parallelism for scalable and elastic stream processing,'' \emph{IEEE
  Transactions on Parallel and Distributed Systems}, vol.~33, no.~12, pp.
  4221--4238, 2022.

\bibitem{trofimov2022substream}
A.~Trofimov, N.~Sokolov, N.~Marshalkin, I.~Kuralenok, and B.~Novikov,
  ``Substream management in distributed streaming dataflows,'' in
  \emph{Proceedings of the 16th ACM International Conference on Distributed and
  Event-Based Systems}, 2022, pp. 55--66.

\bibitem{WikiAtomicEdits}
M.~Faruqui, E.~Pavlick, I.~Tenney, and D.~Das, ``{WikiAtomicEdits: A
  Multilingual Corpus of Wikipedia Edits for Modeling Language and
  Discourse},'' in \emph{Proc. of EMNLP}, 2018.

\bibitem{mohamed2018detection}
I.~S. Mohamed, A.~Capitanelli, F.~Mastrogiovanni, S.~Rovetta, and R.~Zaccaria,
  ``Detection, localisation and tracking of pallets using machine learning
  techniques and 2d range data,'' \emph{arXiv preprint arXiv:1803.11254}, 2018.

\bibitem{mohamed20192d}
------, ``A 2d laser rangefinder scans dataset of standard eur pallets,''
  \emph{Data in Brief}, p. 103837, 2019.

\bibitem{OdroidXU42016a}
\BIBentryALTinterwordspacing
HardKernel, ``{Odroid-XU4},'' 2020. [Online]. Available:
  \url{http://www.hardkernel.com}
\BIBentrySTDinterwordspacing

\bibitem{fuEdgeWiseBetterStream}
X.~Fu, T.~Ghaffar, J.~C. Davis, and D.~Lee, ``Edgewise: A better stream
  processing engine for the edge,'' in \emph{{{USENIX}} Annual Technical
  Conference ({{ATC}}) 19}.\hskip 1em plus 0.5em minus 0.4em\relax {WA, USA}:
  USENIX, Jul. 2019, pp. 929--946.

\bibitem{palyvos-giannasHarenFrameworkAdHoc2019}
D.~{Palyvos-Giannas}, V.~Gulisano, and M.~Papatriantafilou, ``Haren: {{A
  Framework}} for {{Ad}}-{{Hoc Thread Scheduling Policies}} for {{Data
  Streaming Applications}},'' in \emph{Proceedings of the 13th {{ACM
  International Conference}} on {{Distributed}} and {{Event}}-Based
  {{Systems}}}, ser. {{DEBS}} '19.\hskip 1em plus 0.5em minus 0.4em\relax
  {Darmstadt, Germany}: {ACM}, 2019, pp. 19--30.

\bibitem{openissue}
``{Apache Flink - Rework streaming iteration flow control},''
  \url{https://issues.apache.org/jira/browse/FLINK-2497?jql=project%20%3D%20FLINK%20AND%20text%20~%20%22loop%20deadlock%22},
  accessed:2023-1-27.

\bibitem{gulisano2017performance}
V.~Gulisano, A.~V. Papadopoulos, Y.~Nikolakopoulos, M.~Papatriantafilou, and
  P.~Tsigas, ``Performance modeling of stream joins,'' in \emph{Proceedings of
  the 11th ACM International Conference on Distributed and Event-based
  Systems}, 2017, pp. 191--202.

\bibitem{dean:CACM:2008:mapreduce}
J.~Dean and S.~Ghemawat, ``Mapreduce: Simplified data processing on large
  clusters,'' \emph{Communications of the ACM}, vol.~51, no.~1, pp. 107--113,
  2008.

\bibitem{abadi2003aurora}
\BIBentryALTinterwordspacing
D.~J. Abadi, D.~Carney, U.~\c{C}etintemel, M.~Cherniack, C.~Convey, S.~Lee,
  M.~Stonebraker, N.~Tatbul, and S.~Zdonik, ``Aurora: A new model and
  architecture for data stream management,'' \emph{The VLDB Journal}, vol.~12,
  no.~2, p. 120–139, 2003. [Online]. Available:
  \url{https://doi.org/10.1007/s00778-003-0095-z}
\BIBentrySTDinterwordspacing

\bibitem{margara2022dataintensive}
\BIBentryALTinterwordspacing
A.~Margara, G.~Cugola, N.~Felicioni, and S.~Cilloni, ``A model and survey of
  distributed data-intensive systems,'' 2022. [Online]. Available:
  \url{https://arxiv.org/abs/2203.10836}
\BIBentrySTDinterwordspacing

\bibitem{zaharia:CACM:2016:spark}
M.~Zaharia, R.~S. Xin, P.~Wendell, T.~Das, M.~Armbrust, A.~Dave, X.~Meng,
  J.~Rosen, S.~Venkataraman, M.~J. Franklin, A.~Ghodsi, J.~Gonzalez,
  S.~Shenker, and I.~Stoica, ``Apache spark: A unified engine for big data
  processing,'' \emph{Communications of the ACM}, vol.~59, no.~11, pp. 56--65,
  2016.

\bibitem{carbone:IEEEB:2015:flink}
P.~Carbone, A.~Katsifodimos, S.~Ewen, V.~Markl, S.~Haridi, and K.~Tzoumas,
  ``Apache flink{\texttrademark}: Stream and batch processing in a single
  engine,'' \emph{IEEE Data Engineering Bulletin}, vol.~38, no.~4, pp. 28--38,
  2015.

\bibitem{toshniwal:SIGMOD:2014:Storm}
A.~Toshniwal, S.~Taneja, A.~Shukla, K.~Ramasamy, J.~M. Patel, S.~Kulkarni,
  J.~Jackson, K.~Gade, M.~Fu, J.~Donham, N.~Bhagat, S.~Mittal, and D.~Ryaboy,
  ``Storm@twitter,'' in \emph{Proc of the Intl Conf on Management of Data},
  ser. SIGMOD '14.\hskip 1em plus 0.5em minus 0.4em\relax ACM, 2014, pp.
  147--156.

\bibitem{kulkarni:SIGMOD:2015:Twitter_Heron}
S.~Kulkarni, N.~Bhagat, M.~Fu, V.~Kedigehalli, C.~Kellogg, S.~Mittal, J.~M.
  Patel, K.~Ramasamy, and S.~Taneja, ``Twitter heron: Stream processing at
  scale,'' in \emph{Proc of the Intl Conf on Management of Data}, ser. SIGMOD
  '15.\hskip 1em plus 0.5em minus 0.4em\relax ACM, 2015, pp. 239--250.

\bibitem{bejeck:2018:KafkaStreams}
B.~Bejeck, \emph{Kafka Streams in Action: Real-time apps and microservices with
  the Kafka Streams API}.\hskip 1em plus 0.5em minus 0.4em\relax Manning, 2018.

\bibitem{noghabi:VLDB:2017:Samza}
S.~A. Noghabi, K.~Paramasivam, Y.~Pan, N.~Ramesh, J.~Bringhurst, I.~Gupta, and
  R.~H. Campbell, ``Samza: Stateful scalable stream processing at linkedin,''
  \emph{Proc of VLDB}, vol.~10, no.~12, p. 1634–1645, 2017.

\bibitem{shute:VLDB:2013:F1}
J.~Shute, R.~Vingralek, B.~Samwel, B.~Handy, C.~Whipkey, E.~Rollins, M.~Oancea,
  K.~Littlefield, D.~Menestrina, S.~Ellner, J.~Cieslewicz, I.~Rae,
  T.~Stancescu, and H.~Apte, ``F1: A distributed sql database that scales,''
  \emph{Proc of VLDB}, vol.~6, no.~11, p. 1068–1079, 2013.

\bibitem{begoli2019one}
E.~Begoli, T.~Akidau, F.~Hueske, J.~Hyde, K.~Knight, and K.~Knowles, ``One sql
  to rule them all-an efficient and syntactically idiomatic approach to
  management of streams and tables,'' in \emph{Proceedings of the 2019
  International Conference on Management of Data}, 2019, pp. 1757--1772.

\bibitem{margara:JWS:2014:streaming}
A.~Margara, J.~Urbani, F.~{van Harmelen}, and H.~Bal, ``Streaming the web:
  Reasoning over dynamic data,'' \emph{Journal of Web Semantics}, vol.~25, pp.
  24--44, 2014.

\bibitem{hirzel2018stream}
M.~Hirzel, G.~Baudart, A.~Bonifati, E.~Della~Valle, S.~Sakr, and
  A.~Akrivi~Vlachou, ``Stream processing languages in the big data era,''
  \emph{ACM Sigmod Record}, vol.~47, no.~2, pp. 29--40, 2018.

\bibitem{soule2010universal}
R.~Soul{\'e}, M.~Hirzel, R.~Grimm, B.~Gedik, H.~Andrade, V.~Kumar, and K.-L.
  Wu, ``A universal calculus for stream processing languages,'' in
  \emph{Programming Languages and Systems: 19th European Symposium on
  Programming, ESOP 2010, Held as Part of the Joint European Conferences on
  Theory and Practice of Software, ETAPS 2010, Paphos, Cyprus, March 20-28,
  2010. Proceedings 19}.\hskip 1em plus 0.5em minus 0.4em\relax Springer, 2010,
  pp. 507--528.

\bibitem{arasu2006cql}
A.~Arasu, S.~Babu, and J.~Widom, ``The cql continuous query language: semantic
  foundations and query execution,'' \emph{The VLDB Journal}, vol.~15, pp.
  121--142, 2006.

\bibitem{botan:VLDB:2010:SECRET}
\BIBentryALTinterwordspacing
I.~Botan, R.~Derakhshan, N.~Dindar, L.~Haas, R.~J. Miller, and N.~Tatbul,
  ``Secret: A model for analysis of the execution semantics of stream
  processing systems,'' \emph{Proc. VLDB Endow.}, vol.~3, no. 1–2, p.
  232–243, 2010. [Online]. Available:
  \url{https://doi.org/10.14778/1920841.1920874}
\BIBentrySTDinterwordspacing

\bibitem{affetti2017defining}
L.~Affetti, R.~Tommasini, A.~Margara, G.~Cugola, and E.~Della~Valle, ``Defining
  the execution semantics of stream processing engines,'' \emph{Journal of Big
  Data}, vol.~4, no.~1, pp. 1--24, 2017.

\bibitem{akidauwatermarks}
T.~Akidau, E.~Begoli, S.~Chernyak, F.~Hueske, K.~Knight, K.~Knowles, D.~Mills,
  and D.~Sotolongo, ``Watermarks in stream processing systems: Semantics and
  comparative analysis of apache flink and google cloud dataflow,''
  \emph{Proceedings of the VLDB Endowment}, no.~3, 2020.

\bibitem{gevay:CSur:2022:iterations}
\BIBentryALTinterwordspacing
G.~E. G\'{e}vay, J.~Soto, and V.~Markl, ``Handling iterations in distributed
  dataflow systems,'' \emph{ACM Comput. Surv.}, vol.~54, no.~9, 2021. [Online].
  Available: \url{https://doi.org/10.1145/3477602}
\BIBentrySTDinterwordspacing

\bibitem{sax:BIRTE:2018:StreamsTables}
\BIBentryALTinterwordspacing
M.~J. Sax, G.~Wang, M.~Weidlich, and J.-C. Freytag, ``Streams and tables: Two
  sides of the same coin,'' in \emph{Proceedings of the International Workshop
  on Real-Time Business Intelligence and Analytics}, ser. BIRTE '18.\hskip 1em
  plus 0.5em minus 0.4em\relax New York, NY, USA: Association for Computing
  Machinery, 2018. [Online]. Available:
  \url{https://doi.org/10.1145/3242153.3242155}
\BIBentrySTDinterwordspacing

\bibitem{fernandez:ATC:2014:MakingStateExplicit}
R.~C. Fernandez, M.~Migliavacca, E.~Kalyvianaki, and P.~Pietzuch, ``Making
  state explicit for imperative big data processing,'' in \emph{Proc of the
  USENIX Annual Technical Conf}, ser. ATC'14.\hskip 1em plus 0.5em minus
  0.4em\relax USENIX Assoc., 2014, p. 49–60.

\bibitem{murray:SOSP:2013:Naiad}
D.~G. Murray, F.~McSherry, R.~Isaacs, M.~Isard, P.~Barham, and M.~Abadi,
  ``Naiad: A timely dataflow system,'' in \emph{Proc of the Symposium on
  Operating Systems Principles}, ser. SOSP '13.\hskip 1em plus 0.5em minus
  0.4em\relax ACM, 2013, p. 439–455.

\bibitem{murray:NSDI:2011:CIEL}
D.~G. Murray, M.~Schwarzkopf, C.~Smowton, S.~Smith, A.~Madhavapeddy, and
  S.~Hand, ``Ciel: A universal execution engine for distributed data-flow
  computing,'' in \emph{Proc of the Conf on Networked Systems Design and
  Implementation}, ser. NSDI'11.\hskip 1em plus 0.5em minus 0.4em\relax USENIX
  Assoc., 2011, p. 113–126.

\bibitem{affetti:JPDC:2020:tspoon}
L.~Affetti, A.~Margara, and G.~Cugola, ``Tspoon: Transactions on a stream
  processor,'' \emph{Journal of Parallel and Distributed Computing}, vol. 140,
  pp. 65--79, 2020.

\bibitem{cetintemel:VLDB:2014:sstore}
U.~Cetintemel, J.~Du, T.~Kraska, S.~Madden, D.~Maier, J.~Meehan, A.~Pavlo,
  M.~Stonebraker, E.~Sutherland, N.~Tatbul \emph{et~al.}, ``S-store: a
  streaming newsql system for big velocity applications,'' \emph{Proc of VLDB},
  vol.~7, no.~13, pp. 1633--1636, 2014.

\bibitem{cugola:CSur:2012:survey}
\BIBentryALTinterwordspacing
G.~Cugola and A.~Margara, ``Processing flows of information: From data stream
  to complex event processing,'' \emph{ACM Comput. Surv.}, vol.~44, no.~3,
  2012. [Online]. Available: \url{https://doi.org/10.1145/2187671.2187677}
\BIBentrySTDinterwordspacing

\bibitem{giatrakos:VLDBJ:2020:CER}
\BIBentryALTinterwordspacing
N.~Giatrakos, E.~Alevizos, A.~Artikis, A.~Deligiannakis, and M.~Garofalakis,
  ``Complex event recognition in the big data era: A survey,'' \emph{The VLDB
  Journal}, vol.~29, no.~1, p. 313–352, jul 2019. [Online]. Available:
  \url{https://doi.org/10.1007/s00778-019-00557-w}
\BIBentrySTDinterwordspacing

\bibitem{grez:TODS:2021:Formal}
\BIBentryALTinterwordspacing
A.~Grez, C.~Riveros, M.~Ugarte, and S.~Vansummeren, ``A formal framework for
  complex event recognition,'' \emph{ACM Trans. Database Syst.}, vol.~46,
  no.~4, dec 2021. [Online]. Available: \url{https://doi.org/10.1145/3485463}
\BIBentrySTDinterwordspacing

\bibitem{shchultz:DEBS:2009:rewriting}
\BIBentryALTinterwordspacing
N.~P. Schultz-M\o{}ller, M.~Migliavacca, and P.~Pietzuch, ``Distributed complex
  event processing with query rewriting,'' in \emph{Proceedings of the Third
  ACM International Conference on Distributed Event-Based Systems}, ser. DEBS
  '09.\hskip 1em plus 0.5em minus 0.4em\relax New York, NY, USA: Association
  for Computing Machinery, 2009. [Online]. Available:
  \url{https://doi.org/10.1145/1619258.1619264}
\BIBentrySTDinterwordspacing

\bibitem{Hirzel2014CSP}
M.~Hirzel, R.~Soul{\'e}, S.~Schneider, B.~Gedik, and R.~Grimm, ``A catalog of
  stream processing optimizations,'' \emph{ACM Computing Surveys (CSUR)},
  vol.~46, no.~4, pp. 1--34, 2014.

\bibitem{guerriero:TOSEM:2021:StreamGen}
\BIBentryALTinterwordspacing
M.~Guerriero, D.~A. Tamburri, and E.~D. Nitto, ``Streamgen: Model-driven
  development of distributed streaming applications,'' \emph{ACM Trans. Softw.
  Eng. Methodol.}, vol.~30, no.~1, 2021. [Online]. Available:
  \url{https://doi.org/10.1145/3408895}
\BIBentrySTDinterwordspacing

\end{thebibliography}

\end{document}